\documentclass[journal,twocolumn,final]{IEEEtran}

\IEEEoverridecommandlockouts

\usepackage{amssymb, amsthm, amsmath, graphicx, cite, color, bm, bbm}
\usepackage[caption=false,font=footnotesize]{subfig}
\usepackage[ruled]{algorithm2e}
\usepackage{psfrag}

\usepackage{tikz,pgfplots}
\pgfplotsset{compat=newest}
\usetikzlibrary{graphs,matrix,positioning,calc,fit,backgrounds,chains}
\usepackage[breakable]{tcolorbox}
\tcbuselibrary{breakable}
\usepackage{tikzscale}

\newtheorem{theorem}{Theorem}
\newtheorem{remark}{Remark}
\newtheorem{corollary}{Corollary}

\newtheorem{prop}{Proposition}
\newtheorem{defi}{Definition}

\newcommand{\FA}{{\sf FA}}
\newcommand{\MD}{{\sf MD}}
\newcommand{\D}{{\sf D}}
\newcommand{\LLR}{{\sf LLR}}
\newcommand{\LR}{{\sf LR}}

\let\oldbrace\{
\def\{{\oldbrace\kern0.5pt}

\def\tr{\mathop{\rm tr}\nolimits}%
\newcommand{\nn}{\nonumber}

\newcommand{\Cc}{\mathcal{C}}

\newcommand{\Ic}{\mathcal{I}}

\newcommand{\Nc}{\mathcal{N}}

\newcommand{\Rc}{\mathcal{R}}
\newcommand{\Sc}{\mathcal{S}}
\newcommand{\Tc}{\mathcal{T}}

\newcommand{\Xc}{\mathcal{X}}
\newcommand{\Yc}{\mathcal{Y}}

\newcommand{\Yv}{\boldsymbol{Y}}
\newcommand{\Zv}{\boldsymbol{Z}}

\newcommand{\xv}{\boldsymbol{x}}
\newcommand{\yv}{\boldsymbol{y}}

\newcommand{\Mh}{{\hat{M}}}

\newcommand{\Sh}{{\hat{S}}}

\newcommand{\mh}{{\hat{m}}}
\newcommand{\sh}{{\hat{s}}}

\newcommand{\Xt}{{\tilde{X}}}
\newcommand{\Yt}{{\tilde{Y}}}
\newcommand{\Zt}{{\tilde{Z}}}

\newcommand{\xt}{{\tilde{x}}}
\newcommand{\yt}{{\tilde{y}}}

\def\eps{\epsilon}

\DeclareMathOperator\E{\sf E}
\let\P\relax
\DeclareMathOperator\P{\sf P}

\DeclareMathOperator*{\argmax}{\arg\max}

\begin{document}
	
\title{Integrated Communication and Binary State Detection Under Unequal Error Constraints}

\author{Daewon Seo and Sung Hoon Lim\\
\thanks{D.~Seo is with the Department of Electrical Engineering and Computer Science, Daegu Gyeongbuk Institute of Science and Technology (DGIST), Daegu 42988, South Korea (e-mail: dwseo@dgist.ac.kr).}
\thanks{Sung Hoon Lim is with the School of Information Sciences, Hallym University, Chuncheon 24252, South Korea (e-mail: shlim@hallym.ac.kr).}
}

\maketitle

\allowdisplaybreaks

\begin{abstract}
	This work considers a problem of integrated sensing and communication (ISAC) in which the goal of sensing is to detect a binary state. Unlike most approaches that minimize the total detection error probability, in our work, we disaggregate the error probability into false alarm and missed detection probabilities and investigate their information-theoretic three-way tradeoff including communication data rate. We consider a broadcast channel that consists of a transmitter, a communication receiver, and a detector where the receiver's and the detector's channels are affected by an unknown binary state. We consider and present results on two different state-dependent models. In the first setting, the state is fixed throughout the entire transmission, for which we fully characterize the optimal three-way tradeoff between the coding rate for communication and the two possibly nonidentical error exponents for sensing in the asymptotic regime. The achievability and converse proofs rely on the analysis of the cumulant-generating function of the log-likelihood ratio. In the second setting, the state changes every symbol in an independently and identically distributed (i.i.d.) manner, for which we characterize the optimal tradeoff region based on the analysis of the receiver operating characteristic (ROC) curves.
\end{abstract}

\section{Introduction}
\subsection{Motivation}
The rapid evolution of wireless communication technologies has created an unprecedented demand for higher data throughput. To meet this need, future wireless communication systems are expected to utilize higher frequencies, such as millimeter waves, which also allow the transmitted signal to be used for radar due to their high resolution. In this context, integrated sensing and communication (ISAC) technology is predicted to be one of the key features of the sixth generation (6G) communication networks, seamlessly merging communication and sensing functionalities into a unified system \cite{Bourdoux2020_2, Liu--Huang--Li--wan--Li--Han--Liu--Du--Tan--Lu--Shen--Colone--Chetty2022_2}. Unlike previous generations that treated these aspects separately, ISAC enables networks to simultaneously transmit data and sense the environment, enhancing situational awareness and resource efficiency \cite{Liu--Masouros--Petropulu--Griffiths--Hanzo2020, Sturm--Wiesbeck2011}. This convergence supports advanced applications like autonomous driving, surveillance, and smart cities by a single transmitted signal.

A key aspect of ISAC from a signal design perspective is that a single transmitted signal must function both as an active sensing signal and a data-bearing signal. These possibly conflicting roles create a new tradeoff between data rate and sensing performance. For example, when a binary multiplicative state is fixed throughout the transmission, reference~\cite{Joudeh--Willems2022} characterized a tradeoff between data rate and the error exponent of total sensing errors. This was later extended to a general binary state \cite{Wu--Joudeh2022} and to multiple states \cite{Chang--Wang--Erdogan--Bloch--2023}. This body of literature indeed focuses on the most indistinguishable state, i.e., the worst state-conditional error, leading to a minimax or equalizer design of ISAC. On the other hand, for a state varying every symbol in an independently and identically distributed (i.i.d.) manner, reference \cite{Ahmadipour--Kobayashi--Wigger--Caire2022} characterized a tradeoff between data rate and general Bayesian distortion. Specializing in error probability, it indeed considers the Bayesian probability of error, and hence, uncommon states are allowed to be highly erroneous as long as common states are detected accurately.

However, detection errors consist of several components; for binary state detection, they include false alarms and missed detections, which often need to be separately controlled. Consider a motivating scenario where an autonomous vehicle communicates with roadside infrastructure and simultaneously detects pedestrians in its path using shared transmitted signals. In this case, there are two types of detection errors \cite{Poor1994, MoulinV2019}: false alarms (also called type I or false positive error), which occur when the system alerts the presence of a pedestrian when there is none, and missed detections (also called type II or false negative error), which occur when the system fails to alert even though pedestrians are present. Clearly, a missed detection could lead to catastrophic consequences whereas a false alarm may only result in a slight reduction in efficiency or unnecessary caution. Despite the critical importance of this issue, only a small portion of the ISAC literature addresses the problem of unequal error protection~\cite{Wu--Joudeh2022, AhmadipourWS2024}. Addressing this gap is essential for the reliable deployment of ISAC systems, especially in safety-critical applications.

\begin{table*}[t]
\centering 
\caption{Summary of related works}
\begin{tabular}{ c | c | c | c}
\label{tab:related_work}
 & State model & Performance metric for sensing &  Scope\\
\hline \hline
\cite{Joudeh--Willems2022} & Fixed & Minimax error exponent & \\
\hline
\cite{Chang--Wang--Erdogan--Bloch--2023} & Fixed & Minimax error exponent & \\
\hline
\cite[Sec.~V]{Wu--Joudeh2022} & Fixed & False alarm and missed detection probabilities & \\
\hline
\cite{AhmadipourWS2024} & Fixed & False alarm and missed detection exponents & Implicit tradeoff characterization \\
\hline
Ours (Sec.~\ref{sec:main results}) & Fixed & False alarm and missed detection exponents & Closed-form tradeoff characterization \\
\hline
\hline
\cite{Zhang--Vedantam--Mitra2011} & i.i.d. & General Bayesian distortion & Non-adpative codes \\
\hline
\cite{Ahmadipour--Kobayashi--Wigger--Caire2022} & i.i.d. & General Bayesian distortion & Adpative codes \\
\hline
Ours (Sec.~\ref{sec:iid_state}) & i.i.d. & False alarm and missed detection probabilities & Non-adpative codes \\
\end{tabular}
\end{table*}

\subsection{Main contribution}
This work explores the problem of the unequal error protection in ISAC from an information-theoretic perspective. In particular, we consider two different settings on states. In the first setting, the state is fixed throughout the entire transmission, for which we fully characterize the optimal three-way tradeoff between the coding rate for communication and the two possibly nonidentical error exponents for sensing in the asymptotic regime. The analysis relies on the cumulant-generating function of the log-likelihood ratio. Unlike previous work~\cite{AhmadipourWS2024}, our result explicitly characterize the performance region and provide a constructive detection scheme. In the second setting, the state changes every symbol in an i.i.d.~manner, for which we characterize the optimal tradeoff region based on the Neyman-Pearson analysis with the receiver operating characteristic (ROC) curves. Additionally, by applying the waterfilling algorithm, we explicitly demonstrate how to evaluate and achieve the optimal performace.

Note that the first problem is a single-state detection problem based on multiple observations, i.e., the detection performance scales with block-length. On the other hand, in the second problem, multiple detections are performed in a symbol-by-symbol manner. The difference in the problem settings suggest two different approaches, the first is an asymptotic analysis that reveals how the detection error scales with block-length, while the second approach reveals the finite error tradeoff in a Neyman-Pearson setting. Overall, our goal is to provide a comprehensive understanding of unequal error protection in ISAC for two common settings in the literature.

\subsection{Related works}
Information-theoretic approaches on dual-functional signal design for both data transmission and sensing were initially explored in~\cite{Sutivong--Chiang--Cover--Kim2005, Kim--Sutivon--Cover2008, Choudhuri--Kim--Mitra2020}, under the name of ``state amplification.'' Here, the channel state varies in an independent and identically distributed (i.i.d.) manner and is assumed to be known to the encoder. In this framework, the encoder embeds the state information into the transmitted codewords, enabling the receiver to perform simultaneous data decoding and state estimation. Subsequently, \cite{Zhang--Vedantam--Mitra2011} considered a scenario where the state information is unavailable at the encoder and showed the tradeoff between communication and sensing. In recent monostatic ISAC systems in which transmitters aim to detect the state from backscattered signals, adaptive signaling can be utilized, e.g.,~\cite{Ahmadipour--Kobayashi--Wigger--Caire2022, LiADMB2024}. Such results are extended to a Markov-state ISAC problem in~\cite{GootyM2024}. In the above studies, two features of channel capacity for data transmission and rate distortion for state detection arise together due to the nature of varying states, and hence, the performance is characterized by the so-called capacity-distortion function. If the distortion metric is Hamming, then it indeed measures the Bayesian probability of state detection errors. Hence, it is biased toward more common states and allows high detection error probabilities for uncommon states. This is further extended in our second formulation in Sec.~\ref{sec:iid_state}, where we introduce unequal error constraints. It enables us to gain a more explicit understanding of the performance from the perspective of detection theory.

Another line of research is motivated by physical changes in the sensing environment that occur much slower than the time scale of communication. It assumes that an unknown state remains fixed throughout the communication, such as an obstacle disrupting the sensing path. In this context, reference \cite{Joudeh--Willems2022} focuses on a binary multiplicative state, which is later extended in \cite{Chang--Wang--Erdogan--Bloch--2023} to multiple states that are not necessarily multiplicative. For these two works, the sensing performance is measured by the worst state-conditional error exponent. Hence, the solution is indeed a minimax or equalizer solution. Reference \cite[Sec.~V]{Wu--Joudeh2022} investigates the rate-exponent region in the Neyman-Pearson formulation, where the error exponent for missed detection is minimized under the constraint that the false alarm probability is bounded above by a given $\alpha \in (0,1)$. However, since the false alarm ``probability'' always vanishes for a fixed state as long as the block length of codes grows without bound and a detection rule is nontrivial, the false alarm constraint is always met. Therefore, the resulting error exponent for missed detection is indeed independent of $\alpha$, i.e., the tradeoff between $\alpha$ and the error exponent is obvious. Under the same setting as ours,~\cite{AhmadipourWS2024} investigates false alarm and missed detection error exponents and characterizes the optimal tradeoff region. Compared to our work,~\cite{AhmadipourWS2024} adopts a converse-centric approach, primarily focusing on a strong converse result. Moreover, since the expression for the rate region is formulated as an optimization problem, further effort is required to explicitly characterize the tradeoff and provide a matching constructive achievable scheme. In contrast, our work adopts a more achievability-centric approach, focusing on explicitly evaluating and providing insights into the tradeoff. Our approach leads to a closed-form result with a matching strategy. Recently, \cite[Thm.~1]{Wu--Joudeh2024} refines the preliminary results on an ISAC problem with unequal detection errors \cite{Wu--Joudeh2022} and presents a result with the same expression to our Thm.~\ref{thm:theorem1}. The main distinction is that our communication channel is state-dependent, while \cite[Thm.~1]{Wu--Joudeh2024} is state-free. Related works are summarized in Tab.~\ref{tab:related_work}.

In addition, studying ISAC for sensing continuous targets, such as amplitude, delay, angle, and Doppler, is also a popular research topic \cite{Xiong--Liu--Cui--Yuan--Han--Caire2023, Liu--Yuan--Masouros--Yuan2020, Liu--Liu--Li--Masouros--Eldar2022, Hua--Han--Xu2023, DongLLX2023, RenPSFQLNX2024}. These studies typically focus on the mean-squared error (MSE) and optimizing the Cram\'{e}r-Rao lower bound or its variants as a performance proxy. Extending the analysis from point-to-point settings to multiuser scenarios, such as multiple-access channels \cite{Kobayashi--Hamad--Kramer--Caire2019}, broadcast channels \cite{Ahmadipour--Kobayashi--Wigger--Caire2022}, and relay channels \cite{Liu--Li--Ong--Yener2024}, also remains an open and active area of research.

In stand-alone classic binary hypothesis testing problems, such unequal error probability constraints are studied in the Neyman-Pearson formulation~\cite{Poor1994, MoulinV2019}. Further, when a state is fixed and observations are i.i.d.~from the state, Hoeffding characterizes the optimal nontrivial tradeoff between the two exponents for the probabilities of false alarm and missed detection, respectively~\cite{Hoeffding1965}. Our work extends these results to the ISAC scenarios.

\subsection{Organization and notation}
The remainder of this paper is organized as follows. We formally define our problem in Sec.~\ref{sec:formulation}. The results for a fixed state and i.i.d.~states are presented respectively in Sec.~\ref{sec:main results} and Sec.~\ref{sec:iid_state}. Sec.~\ref{sec:discussion} concludes the paper.

This paper has been partially presented at the 22nd International Symposium on Modeling and Optimization in Mobile, Ad hoc, and Wireless Networks (WiOpt 2024) \cite{Seo--Lim_wiopt2024}, where only a fixed-state scenario with a state-independent communication channel is considered. This paper extends it to a more general scenario with a state-dependent communication channel in Sec.~\ref{sec:main results}. Furthermore, the case of an i.i.d.~varying state is additionally explored in Sec.~\ref{sec:iid_state}.

A notable departure from standard notations in this paper is that we use $D(p(Y|x)\| q(Y|x)):= D(p_{Y|X=x}\|q_{Y|X=x})$ to clearly state KL divergences with distributions conditioned on a specific realization of channel input. Also, we use $I(X;\Yt) = I(p_X, p_{\Yt|X})$ interchangeably for the mutual information to emphasize its dependency. The type of sequence $x^n$ is denoted by $\pi(x^n)$. If the $i$-th element of a length $n$ vector is removed, it is denoted by $x_{-i} = (x_1, \ldots, x_{i-1}, x_{i+1}, \ldots, x_n)$.

\section{Problem Statement} \label{sec:formulation}
\begin{figure}[!t]
	\begin{center}
		\resizebox{!}{10.0em}{
			\begin{tikzpicture}[font=\large,node distance=.6cm and 1cm, start chain]
     \tikzstyle{rect}=[draw=black, 
                   rectangle, 
                   text opacity=1,
                   minimum width=50pt, 
                   minimum height = 25pt, 
                   align=center]
  \node[rect] (encoder) {Encoder};
  \node[rect,above right=0.4cm and 1.6cm of encoder] (com_chn) {$p_{\tilde{Y}|X,S}$};
  \node[rect,below right=0.4cm and 1.6cm of encoder] (qcd_chn) {$p_{Y|X, S}$};
  \node[rect, right=of com_chn] (decoder) {Decoder};
  \node[rect, right=of qcd_chn] (detector) {Detector};
\draw[-stealth] (encoder.east) -- node[above] {$X^n$} +(1,0) |- (com_chn.west);
\draw[-stealth] (encoder.east) -- +(1,0) |- (qcd_chn.west);
\draw[-stealth] (com_chn.east) -- node[above] {$\tilde{Y}^n$}(decoder.west);
\draw[-stealth] (qcd_chn.east) -- node[above] {$Y^n$}(detector.west);
\draw[stealth-] (encoder.west) --+ (-5mm,0) node[left] {$M$};
\draw[-stealth] (decoder.east) --+ (5mm,0) node[right] {$\hat{M}$};
\draw[-stealth] (detector.east) --+ (5mm,0) node[right] {$\hat{S}$ or $\hat{S}^n$};
\draw[stealth-] (detector.south) --+ (0,-5mm) node[below] {$M$};
\end{tikzpicture}}
	\end{center}
	\caption{Channel model for the joint communication and binary state detection problem.}
    \label{fig:model1}
\end{figure}
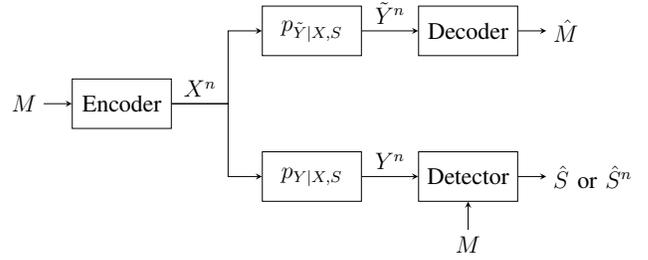

We consider a discrete memoryless\footnote{As can be seen in \eqref{eq:DMC_pmf}, the channel is memoryless with respect to the channel input conditioned on the states.} state-dependent three-node network, depicted in Fig.~\ref{fig:model1} that consists of a transmitter, a receiver, and a detector. This serves as a canonical model for bistatic ISAC scenarios where side information is available at the detector. For example, it is an abstract model for the case in which a base station transmits to a receiver while a second base station acts as a sensor with a backhaul wireline connection to the transmitting basestation. The channel model also serves as a model for the case the encoder and detector are coupled in a single device (a monostatic ISAC system) that does not use feedback adaptive codes (e.g., open-loop coding \cite{Chang--Wang--Erdogan--Bloch--2023}). In the latter case, $p_{Y|X,S}$ models the echo channel.

\subsection{Statement for a Fixed State}
When a state is fixed throughout the entire transmission, e.g., \cite{Joudeh--Willems2022, Wu--Joudeh2022, Chang--Wang--Erdogan--Bloch--2023}, the distribution follows
\begin{align}
	p(\yt^n, y^n|x^n, s) = \prod_{i=1}^np(\yt_i|x_i, s)p(y_i|x_i, s) \label{eq:DMC_pmf}
\end{align}
where $\yt^n \in \tilde{\Yc}^n$ is the observation sequence at the receiver, $y^n \in \Yc^n$ is the observation sequence at the detector, and $s\in\Sc=\{0,1\}$ is an unknown state that is fixed throughout the transmission. For simplicity, we will often denote $p_s(y|x) := p(y|x,s)$. 

The transmitter wishes to send a message $m\in[1:2^{nR}]$ to the receiver while simultaneously enabling the detector to estimate the unknown state $S$. For the detector, we assume that the message $m$ is given as side information. Formally, a $(2^{nR}, n)$ code for the joint communication and detection problem consists of
\begin{itemize}
	\item a message set $m\in[1:2^{nR}]$, 
	\item an encoder that assigns a sequence $x^n(m) \in \Xc^n$ to each message $m\in[1:2^{nR}]$, 
	\item a decoder that assigns a message estimate $\mh(\yt^n)$ to each observation sequence $\yt^n$, and
	\item a detector that assigns a state estimate $\sh(y^n, x^n(m))\in\Sc$ to each detector's observation sequence $y^n$ and message side information $m$.
\end{itemize}
We denote the codebook by $\Cc^{(n)} = \{x^n(m), m\in[1:2^{nR}]\}$, and suppose alphabet spaces are finite, i.e., $|\Xc|, |\Yc|, |\tilde{\Yc}| < \infty$. Gaussian cases are treated by quantization argument, e.g., \cite{El-Gamal--Kim2011}.

The performance metrics are defined as follows. The probability of error for communication is defined as $P_e^{(n)} = \P(M\neq \Mh)$, i.e., an average probability of error. We define two performance metrics for detection, the false alarm and missed detection probabilities as follows.

\begin{defi}
For a given codeword $x^n$, the false alarm and missed detection probabilities are respectively defined as
\begin{align*}
	P^{(n)}_\FA(x^n) &= \P[ \hat{S}(Y^n, x^n) = 1 | X^n(M) = x^n, S = 0 ], \\
	P^{(n)}_\MD(x^n) &= \P[ \hat{S}(Y^n, x^n) = 0 | X^n(M) = x^n, S = 1 ].
\end{align*}
\end{defi}

Since we have $n$ noisy observations for a given state, error probabilities typically decay exponentially fast in $n$. To this end, we say that a rate-exponent tuple $(R, E_{\FA}, E_{\MD})$ is achievable if there exists a sequence of $(2^{nR}, n)$ codes such that $\lim_{n\to\infty} P^{(n)}_e = 0$ and
\begin{align*}
	\liminf_{n \to \infty} \min_{x^n \in \Cc^{(n)}} -\frac{1}{n} \log P^{(n)}_\FA(x^n)  &\ge E_\FA, \\
	\liminf_{n \to \infty} \min_{x^n \in \Cc^{(n)}} -\frac{1}{n} \log P^{(n)}_\MD(x^n) &\ge E_\MD,
\end{align*}
simultaneously. The optimal region is the set of all achievable tuples. Then, the rate-exponent region $\Rc$ is the closure of the set of all achievable tuples $(R, E_\FA, E_\MD)$, that is 
\begin{align*}
	\Rc := \text{cl}\{(R,E_\FA,E_\MD): (R,E_\FA,E_\MD) \text{ is achievable} \}.
\end{align*}
Therefore, our goal is to find the optimal tradeoff between the data rate $R$ and error exponents $E_\FA, E_\MD$.

In this work, we are particularly interested in the regime where both $E_\MD > 0$ and $E_\FA > 0$, since otherwise the problem becomes trivial. For instance, for the case $E_\FA = 0$, the detector can simply declare $\hat{S}=1$ always regardless of observations. It gives $(E_\FA, E_\MD) = (0, \infty)$. Similarly, it is clear that we can also achieve $(E_\FA, E_\MD) = (\infty, 0)$. Therefore, we focus only on the nontrivial regime where both $E_\FA, E_\MD$ are strictly positive.

We assume the following technical condition that is necessary for our converse proof: There exists a constant $C > 0$ such that for any $x \in \Xc$,
\begin{align}
	\text{Var}\left( \log \frac{p_1(Y|x)}{p_0(Y|x)} \right) < C. \label{eq:assupmt1}
\end{align}
Note that the condition is mild; the class of channels satisfying the condition includes discrete channels with finite alphabet spaces and an additive white Gaussian noise (AWGN) channel.

\subsection{Statement for i.i.d.~States}
When the state varies in an i.i.d.~manner, e.g., \cite{Ahmadipour--Kobayashi--Wigger--Caire2022}, the distribution follows
\begin{align*}
	p(s^n, \yt^n, y^n|x^n) = \prod_{i=1}^n p(s_i) p(\yt_i|x_i, s_i)p(y_i|x_i, s_i)
\end{align*}
where $s^n \in \Sc^n$ is an unknown sequence of states i.i.d.~distributed according to $p(s)$, $\yt^n \in \tilde{\Yc}^n$ is the observation sequence at the communication receiver, and $y^n \in \Yc^n$ is the observation sequence at the state detector. We assume that $p(s)$ is known to the transmitter and the communication receiver.\footnote{Since we focus on state-conditional errors, providing the detector with knowledge of $p(s)$ does not improve error probabilities.}

Similarly to the fixed case, the transmitter sends a message $m \in [1:2^{nR}]$, and for the detector, we assume that the message $m$ is given as side information. The only difference from the fixed case is that for each time $i$, 
\begin{itemize}
	\item a detector that (possibly randomly) assigns a state estimate $\sh_i(y^n, x^n(m)) \in \Sc$ to each detector's observation sequence $y^n$ and message side information $m$.
\end{itemize}

The performance metrics should also be slightly modified as follows. The probability of error for communication is defined as $P_e^{(n)} = \P(M\neq \Mh)$. For detection, the symbolwise false alarm and missed detection probabilities for the $i$-th state are defined as follows.

\begin{defi}
    For a given codeword $x^n$, the $i$-th state false alarm and missed detection probabilities are respectively defined as
    \begin{align*}
    	P^{(n)}_{\FA,i}(x^n) &= \P[ \hat{S}_i(Y^n, x^n) = 1 | X^n(M) = x^n, S_i = 0 ], \\
    	P^{(n)}_{\MD,i}(x^n) &= \P[ \hat{S}_i(Y^n, x^n) = 0 | X^n(M) = x^n, S_i = 1 ].
    \end{align*}
\end{defi}

Then, the probability of correct detection for the $i$-th state is defined as $P^{(n)}_{\D, i} (x^n) = 1 - P^{(n)}_{\MD,i}(x^n)$. 

We say that for $\alpha, \beta \in (0,1)$, a rate-probability tuple $(R, \alpha, \beta)$ is achievable if there exists a sequence of $(2^{nR}, n)$ codes such that $\lim_{n\to\infty} P^{(n)}_e = 0$ and
\begin{align}
	\liminf_{n \to \infty} \min_{x^n \in \Cc^{(n)}} \frac{1}{n} \sum_{i=1}^n P^{(n)}_{\D,i}(x^n) &\ge \beta, \label{eq:iid_PD_constraint} \\
	\limsup_{n \to \infty} \max_{x^n \in \Cc^{(n)}} \frac{1}{n} \sum_{i=1}^n P^{(n)}_{\FA,i}(x^n)  &\le \alpha \label{eq:iid_FA_constraint}
\end{align}
simultaneously. The optimal rate-probability region is the set of all achievable tuples. Then, the rate-probability region $\Rc$ is the closure of the set of all achievable pairs $(R, \alpha, \beta)$, that is 
\begin{align*}
	\Rc := \text{cl}\{(R, \alpha, \beta): (R, \alpha, \beta) \text{ is achievable} \}.
\end{align*}
It is also convenient to consider a cross section of $\Rc$ at a given $\alpha \in (0,1)$ (cf.~Neyman-Pearson formulation \cite{Poor1994, MoulinV2019}):
\begin{align*}
	\Rc(\alpha) := \text{cl}\{(R, \beta): (R, \alpha, \beta) \text{ is achievable} \}.
\end{align*}

\section{Tradeoff for a fixed state}\label{sec:main results}

The following theorem characterizes the full tradeoff region between the communication rate and two error exponents. Proofs and numerical examples are presented in the next subsections.
\begin{theorem}\label{thm:theorem1}
	The rate-exponent region $\Rc$ is the set of tuples $(R, E_\FA, E_\MD)$ such that
	\begin{align*}
		R &\le \min_{s=0,1} I(p_X, p_{\Yt|X,s}), \\
		E_\FA &\le D( p_u(Y|X) \| p_0(Y|X) | p_X ), \\
		E_\MD &\le D( p_u(Y|X) \| p_1(Y|X) | p_X ),
	\end{align*}
	for some $p_X(x)$ and $u \in (0,1)$ where
	\begin{align*}
		p_u(y|x) = \frac{p_0^{1-u}(y|x)p_1^u(y|x)}{\sum_{y'}p_0^{1-u}(y'|x) p_1^u(y'|x)}.
	\end{align*}
    Moreover, the detection performance corresponding to each $u$ can be obtained by the log-likelihood ratio test such that
    \begin{align*}
    	\hat{S} = \begin{cases}
    		0 & \text{if } \frac{1}{n} \log \frac{p_1(y^n|x^n)}{p_0(y^n|x^n)} < \tau, \\
    		1 & \text{if } \frac{1}{n} \log \frac{p_1(y^n|x^n)}{p_0(y^n|x^n)} \ge \tau,
    	\end{cases}
    \end{align*}
    where $\tau = D(p_{u} \| p_0 | p_X) - D(p_{u} \| p_1 | p_X)$.    
\end{theorem}
In the theorem, $\Rc$ and the optimal detection strategy both are explicitly provided with the parameters $p_X$ and $u$. This is a key distinction from previous work~\cite{AhmadipourWS2024}, where the primary focus is on the strong converse result, and the region and strategy are only given implicitly in the form of an optimization problem. In this sense, our work is more constructive. Several additional remarks follow.

It should be also noted that as usual in information-theoretic works~\cite{CoverT1991}, $p_X$ represents the input distribution of channels, which controls the amount of information in the codes. On the other hand, as we will see, the variable $u$ is a parameter that controls the threshold of a log-likelihood ratio test, which in turn controls the tradeoff between two error exponents. Thus, for a fixed rate and code, the tradeoff between the two error exponents is determined by the detector. For example, assume that the detector receives a sequence of observations $y^n$, generated from a codeword. By adjusting $u$, the detector adjusts the threshold of the log-likelihood test {\em on this single sequence of observations}, and then, the detector can achieve multiple error exponent boundary points. This contrasts with typical rate regions in information theory, where each operating point is obtained by a different codebook, i.e., the rate region is determined by the encoder. In these cases, the receivers perform maximum likelihood (ML) decoding for each specific codebook.

\begin{remark}
Thm.~\ref{thm:theorem1} extends \cite{Joudeh--Willems2022} and a binary version of \cite[Thm.~1]{Chang--Wang--Erdogan--Bloch--2023} to unequal error protection. This can be seen if $u^*$ is taken so that $D(p_{u^*} \| p_0 | p_X) = D(p_{u^*} \| p_1 | p_X)$. This will be revisited at the end of the achievability proof and Cor.~\ref{cor:willems_example} in numerical examples.    
\end{remark}
\begin{remark}
When the communication channel is state independent, the data rate bound reduces to $R \le I(p_X, p_{\Yt|X})$ as shown in our conference version~\cite{Seo--Lim_wiopt2024}.
\end{remark}
\begin{remark}
For a baseline comparison, we consider a na\"{i}ve time-sharing policy that switches between two strategies that attain the extreme corner points. In particular, for $\alpha \in (0,1)$ fraction of the time we use a stand-alone communication-optimal scheme and in the remaining fraction of the time a stand-alone detection-optimal scheme is utilized. Assuming that each strategy ignores the signal outside of the allocated time, the corresponding achievable tuples are as follows:
\begin{align*}
    R &\le \alpha \min_{s=0,1} I(p_X, p_{\Yt|X,s}), \\
    E_\FA &\le (1-\alpha) D( p_u(Y|X) \| p_0(Y|X) | p_X ), \\
    E_\MD &\le (1-\alpha) D( p_u(Y|X) \| p_1(Y|X) | p_X ),
\end{align*}
for some $p_X$ and $u \in (0,1)$.
\end{remark}

\subsection{Preliminaries} \label{subsec:prelim}
We provide some preliminaries that are required to prove Thm.~\ref{thm:theorem1}. We denote the expectation over $p_s(y|x), s=0,1$ by $\E_{s}[\cdot]$, and hence, $\E_0, \E_1$ are the expectations over $p_0(y|x)$ and $p_1(y|x)$, respectively. Similarly, $\E_u$ is the expectation over $p_u(y|x)$, which will be defined later. For simplicity, denote the likelihood ratios by
\begin{align*}
	\LR(y|x) &:= \frac{p_1(y|x)}{p_0(y|x)}, ~~ \LR(y^n|x^n) := \frac{p_1(y^n|x^n)}{p_0(y^n|x^n)},
\end{align*}
and the log-likelihood ratios by $\LLR(\cdot|\cdot) := \log\LR(\cdot|\cdot)$.

For $u \in \mathbb{R}$, $s \in \{0,1\}$, and $x^n$, define the normalized cumulant-generating function (CGF) of the $n$-length log-likelihood ratio as
\begin{align*}
	\kappa_s(u|x^n) &:=  \frac{1}{n} \log \E_s \left[ \exp\left( u \cdot \LLR(Y^n|x^n) \right) \right].
\end{align*}
Note that $\kappa_0$ can be further simplified as
\begin{align*}
	\kappa_0(u|x^n) &:= \frac{1}{n} \log \E_0 \left[ \exp\left( u \log \frac{p_1(Y^n | x^n)}{p_0(Y^n | x^n)} \right) \right] \\
	&= \frac{1}{n} \log \E_0 \left[ \frac{p_1^u(Y^n | x^n)}{p_0^u(Y^n | x^n)} \right] \\
	&\stackrel{(a)}{=} \frac{1}{n} \log \prod_{i=1}^n \E_0 \left[ \frac{p_1^u(Y_i | x_i)}{p_0^u(Y_i | x_i)} \right] \\
	&\stackrel{(b)}{=} \frac{1}{n} \log \prod_{x \in \mathcal{X}} \left( \E_0 \left[ \frac{p_1^u(Y | x)}{p_0^u(Y | x)} \right] \right)^{n_x} \\
	&= \sum_{x \in \mathcal{X}} \frac{n_x}{n} \log  \left( \E_0 \left[ \frac{p_1^u(Y | x)}{p_0^u(Y | x)} \right] \right) \\
	&\stackrel{(c)}{=} \sum_{x \in \mathcal{X}} \hat{p}_X(x) \log  \left( \E_0 \left[ \frac{p_1^u(Y | x)}{p_0^u(Y | x)} \right] \right)\\
	&= \sum_{x \in \mathcal{X}} \hat{p}_X(x) \log  \left( \sum_y p_0^{1-u}(y | x) p_1^u(y | x) \right), 
\end{align*}
where step $(a)$ follows since the observations are conditionally independent, step $(b)$ follows if $n_x$ is defined as the number of occurrence of $x$ in $x^n$, and step $(c)$ follows if $\hat{p}_X$ is the empirical pmf of $x^n$. Note that the negative version of the last expression is a $\hat{p}_X$-weighted sum of Chernoff divergences, which also appears in \cite{Joudeh--Willems2022, Chang--Wang--Erdogan--Bloch--2023} to characterize error exponents.

Similarly, for $v \in \mathbb{R}$,
\begin{align*}
	\kappa_1(v|x^n) &:= \frac{1}{n} \log \E_1 \left[ \exp\left( v \log \frac{p_1(Y_1^n | x_1^n)}{p_0(Y_1^n | x_1^n)} \right) \right] \\
	&= \sum_{x \in \mathcal{X}} \hat{p}_X(x) \log  \left( \E_1 \left[ \frac{p_1^v(Y | x)}{p_0^v(Y | x)} \right] \right)\\
	&= \sum_{x \in \mathcal{X}} \hat{p}_X(x) \log  \left( \sum_y p_0^{-v}(y | x) p_1^{v+1}(y | x) \right).
\end{align*}
It should be noted that
\begin{align}
	\kappa_0(u|x^n) = \kappa_1(v|x^n), \quad \text{ if } u=v+1.\label{eq:prop1}
\end{align}

\noindent\textbf{Property of $\kappa_0$:} Consider a single-letter CGF
\begin{align*}
	\kappa_0(u|x) := \log \E_0 \left[ \exp\left( u \log \frac{p_1(Y | x)}{p_0(Y | x)} \right) \right].
\end{align*}
Then, for every fixed $x$, it is a $U$-shape convex curve crossing zero at $u=0,1$~\cite[Chap.~7]{MoulinV2019}. The normalized multi-letter CGF $\kappa_0(u|x^n)$ is a $\hat{p}_X$-weighted convex combination of the single-letter CGFs, so inherits the properties of $\kappa_0(u|x)$. Let $\kappa_0'(u|x^n)$ be the derivative of $\kappa_0(u|x^n)$ with respect to $u$. Then, for every fixed $x^n$,
\begin{enumerate}
	\item $\kappa_0(u|x^n)$ is strictly convex in $u$ if $p_0 \ne p_1$,
	\item $\kappa_0(0|x^n) = \kappa_0(1|x^n) = 0$,
	\item $\kappa_0'(0|x^n) = -D(p_0(Y|x) \| p_1(Y|x)|\hat{p}_X)$, 
	\item $\kappa_0'(1|x^n) = D(p_1(Y|x) \| p_0(Y|x)|\hat{p}_X)$,
	\item $\kappa_0'(u|x^n)$ satisfies that for $0 \le u \le 1$,
 \begin{align*}
		-D(p_0 \| p_1 | \hat{p}_X) \le \kappa_0'(u|x^n) \le D(p_1 \| p_0 | \hat{p}_X).
	\end{align*}
\end{enumerate}

\subsection{Achievability Proof}
In the achievability proof, we use constant composition codes (CCCs) originally developed in~\cite{Csiszar--Korner2011}. Since our analysis of communication performance relies on the compound channel results from~\cite{Csiszar--Korner2011}, we will only provide a brief summary here, focusing primarily on detection error exponents.

Fix an input type $p_X$ of length $n$, and let $\Tc_{p_X}^n$ be the set of sequences in $\Xc^n$ with composition $p_X$. Then, by the compound channel result~\cite[Cor.~10.10]{Csiszar--Korner2011}, for any $p_X$, the rate $R$ is achievable by some constant composition code $\Cc^{(n)}$ consisting of codewords $x^n(m)\in \Tc_{p_X}^n$ if
\begin{align*}
	R < \min_{s=0,1} I(p_X, p_{\Yt|X,s})
\end{align*}
as $n\to\infty$.

Next, we analyze the performance of the CCC for detection error exponents. Recall that for binary detection problems, it is optimal to perform a log-likelihood ratio test with some proper threshold $\tau \in \mathbb{R}$~\cite{MoulinV2019, Poor1994}. We first consider the case when
\begin{align*}
	-D(p_0 \| p_1 | p_X) < \tau < D(p_1 \| p_0 | p_X),
\end{align*}
which leads us to a nontrivial tradeoff; other trivial cases will be discussed shortly.

Consider an arbitrary codeword $x^n \in \Cc^{(n)}$ from the CCC, and define a decision random variable indicated by the log-likelihood ratio test
\begin{align*}
	\hat{S} = \begin{cases}
		0 & \text{if } \frac{1}{n} \LLR(Y^n | x^n) < \tau, \\
		1 & \text{if } \frac{1}{n} \LLR(Y^n | x^n) \ge \tau.
	\end{cases}
\end{align*}
Then, the false alarm probability can be bounded by the Chernoff bounding technique: For any $u > 0$,
\begin{align}
	P_\FA(x^n) &= \P \left( \frac{1}{n} \LLR(Y^n | x^n) \ge \tau \Big| S=0 \right) \nn \\
	&= \P\left( \exp \left( u \log \frac{p_1(Y^n|x^n)}{p_0(Y^n|x^n)} \right) \ge \exp(n u \tau) \Big| S=0 \right) \nn \\
	&\le \frac{ \E_0 \left[ \exp \left( u\log \frac{p_1(Y^n|x^n)}{p_0(Y^n|x^n)} \right) \right] }{\exp(n u \tau)} \nn \\
	&= \exp( -n(u\tau - \kappa_0 (u|x^n)) ), \label{eq:FA_exp_ub}
\end{align}
where the last equality follows from the definition of $\kappa_0$. Similarly, for any $v < 0$,
\begin{align}
	P_\MD(x^n) &= \P\left( \frac{1}{n} \LLR(Y^n | x^n) < \tau \Big| S=1 \right) \nn \\
	&\le \exp( -n(v\tau - \kappa_1 (v|x^n)) ) \nn \\
	&= \exp( -n( (u-1)\tau - \kappa_0 (u|x^n)) ), \label{eq:MD_exp_ub}
\end{align}
where the last equality holds by choosing $v = u-1$ and the relation given in \eqref{eq:prop1}.

Note that the upper bounds in \eqref{eq:FA_exp_ub} and \eqref{eq:MD_exp_ub} are both minimized simultaneously by the same $u^*$ such that $\kappa_0'(u^*|x^n) = \tau$. To further manipulate expressions and obtain the claimed bounds, we use the (multi-letter) geometric mixture distribution $p_u$ in the following form: For some $u \in (0,1)$,
\begin{align}
	p_u(y^n|x^n) &= \frac{ p_0^{1-u}(y^n|x^n) p_1^{u}(y^n|x^n) }{ \sum_{(y')^n} p_0^{1-u}((y')^n|x^n) p_1^{u}((y')^n|x^n) } \label{eq:geo_mix0} \\
	&= \frac{ p_0(y^n|x^n) \LR^u(y^n|x^n) }{ \E_0[ \LR^u(Y^n|x^n) ] } \label{eq:geo_mix} \\
	&= \frac{ p_0(y^n|x^n) \LR^u(y^n|x^n) }{ \exp( n\kappa_0(u|x^n) ) },\label{eq:geo_mix2}
\end{align}
where
\begin{align*}
    \LR^u(Y^n|x^n) = \left( \LR(Y^n|x^n) \right)^u = \left( \frac{p_1(Y^n|x^n)}{p_0(Y^n|x^n)} \right)^u.
\end{align*}
Then, we have from \eqref{eq:geo_mix2} that
\begin{align}
	&D( p_u(Y^n|x^n) \| p_0(Y^n|x^n) ) \nn \\
	&= \E_{u} \left[ \log \frac{p_u(Y^n|x^n)}{p_0(Y^n|x^n)} \right] \nn \\
	&= u \E_{u}[ \log \LR(Y^n|x^n) ] - n \kappa_0(u|x^n). \label{eq:KLD_mixture1}
\end{align}
For $\E_u[\log \LR(Y^n|x^n)]$ term, we can change the underlying probability measure using \eqref{eq:geo_mix} as follows:
\begin{align*}
	&\E_{u}[ \log \LR(Y^n|x^n) ]\\
	&= \sum_{y^n} p_u(y^n|x^n) \log \LR(y^n|x^n) \\
	&= \sum_{y^n} \frac{ p_0(y^n|x^n) \LR^u(y^n|x^n) }{ \E_0[ \LR^u(Y^n|x^n) ] } \log \LR(y^n|x^n) \\
	&= \frac{ \E_0[\LR^u(Y^n|x^n)  \log \LR(Y^n|x^n) ] }{ \E_0[\LR^u(Y^n|x^n)] } \\
	&= \frac{d}{du} \log \E_0[ \LR^u(Y^n|x^n) ] = n \kappa_0'(u|x^n),
\end{align*}
which gives the relation
\begin{align*}
	\frac{1}{n} D( p_u(Y^n|x^n) \| p_0(Y^n|x^n) ) = u \kappa_0'(u|x^n) - \kappa_0(u|x^n).
\end{align*}
The left side can also be written in a single-letter form by using the additive property of the KL divergence under product distributions.
\begin{align*}
	&D \left( p_u(Y^n|x^n) \| p_0(Y^n|x^n) \right) \\
	&= D \left( \prod_{i=1}^n p_u(Y_i|x_i) \Bigg\| \prod_{i=1}^n p_0(Y_i|x_i) \right) \\
	&= \sum_{i=1}^n D( p_u(Y_i|x_i) \| p_0(Y_i|x_i) ) \\
	&= n D( p_u(Y|X) \| p_0(Y|X) | p_X ),
\end{align*}
which results in 
\begin{align*}
	D( p_u(Y|X) \| p_0(Y|X) | p_X ) = u \kappa_0'(u|x^n) - \kappa_0(u|x^n).
\end{align*}
Therefore, at the optimal $u^*$ such that $\tau = \kappa_0'(u^*|x^n)$,
\begin{align}
	&D( p_{u^*}(Y|X) \| p_0(Y|X) | p_X) \nn \\
	&= u^* \kappa_0'(u^*|x^n) - \kappa_0(u^*|x^n) \label{eq:KLD_p0} \\
	&= \max_u (u\tau - \kappa_0 (u|x^n) ). \nn
\end{align}	
Since $\kappa_0 (u|x^n)$ depends only on the empirical distribution of $x^n$, we can alternatively write $\kappa_0$ as a function of its type $\kappa_0 (u|p_X)$ for any codeword in $\Cc^{(n)}$. In other words, we have a universal bound that holds for all $x^n \in \Cc^{(n)}$ as $\Cc^{(n)}$ is a CCC. Hence, \eqref{eq:FA_exp_ub} can be restated that for any $x^n \in \Cc^{(n)}$,
\begin{align*}
	P_\FA(x^n) &\le \exp( -n \cdot \max_{u \in (0,1)} (u\tau - \kappa_0 (u|p_X)) ) \\
	&= \exp( -n D( p_{u^*}(Y|X) \| p_0(Y|X) | p_X) ).
\end{align*}
Similarly, to bound the error exponent for $P_\MD(x^n)$, it also holds that 
\begin{align}
	&D( p_{u^*}(Y|X) \| p_1(Y|X) | p_X) \nn \\
	&= (u^*-1) \kappa_0'(u^*) - \kappa_0(u^*|x^n) \label{eq:KLD_p1} \\
	&= \max_u ( (u-1)\tau - \kappa_0 (u|x^n) ). \nn
\end{align}
Combined with \eqref{eq:MD_exp_ub}, we have the bound that for any $x^n \in \Cc^{(n)}$,
\begin{align*}
	P_\MD(x^n) \le \exp( -n D( p_{u^*}(Y|X) \| p_1(Y|X) | p_X) ).
\end{align*}

From the strict convexity of $\kappa_0$, $\kappa_0'(u)$ is a monotonically increasing function, implying that $\tau$ and $u^*$ are one-to-one. Therefore, for any given $\tau$ such that
\begin{align*}
	-D(p_0 \| p_1 | p_X) < \tau < D(p_1 \| p_0 | p_X),
\end{align*}
there exists a $u^*$ satisfying $\tau = \kappa_0'(u^*)$.

When $\tau$ does not belong to the above range, the problem becomes trivial. For example, suppose that $\tau \le -D(p_0(Y|X)\|p_1(Y|X)|p_X)$. In this case, the false alarm probability converges to $1$ since
\begin{align*}
	\E_0\left[ \frac{1}{n} \LLR(Y^n|x^n) \right] &= \frac{1}{n} \E_0\left[ \log \frac{p_1(Y^n|x^n)}{p_0(Y^n|x^n)} \right] \\
	&= \frac{1}{n} \sum_{i=1}^n \E_0\left[ \log \frac{p_1(Y_i|x_i)}{p_0(Y_i|x_i)} \right] \\
	&= - \frac{1}{n} \sum_{i=1}^n D( p_0(Y_i|x_i) \| p_1(Y_i|x_i) ) \\
	&= -D( p_0(Y|X) \| p_1(Y|X) | p_X ),
\end{align*}
and thus, the log-likelihood ratio is greater than $\tau$ almost surely as $n\to\infty$, i.e., $P_\FA = 1$, which implies that $E_\FA = 0$. Decreasing $\tau$ further decreases $P_\MD$ (i.e., increases $E_\MD$) at no expense of $E_\FA$; e.g., setting $\tau = -\infty$ achieves $(E_\FA, E_\MD) = (0, \infty)$. The case for $\tau \ge D(p_1(Y|X)\|p_0(Y|X)|p_X)$ can be addressed similarly.

As mentioned, $u^*$ is uniquely determined by $\tau$ through the relationship $\tau = \kappa_0'(u^*|x^n)$. Further based on \eqref{eq:KLD_p0} and \eqref{eq:KLD_p1}, the threshold of the detection rule $\tau$ can be obtained by 
\begin{align}
    \tau = D(p_{u^*} \| p_0 | p_X) - D(p_{u^*} \| p_1 | p_X). \label{eq:tau_relation}
\end{align}
In other words, from two performance bounds on exponents (which later will be shown to be tight in converse), the decision threshold can be determined. Before concluding the proof, the following remark can be made.

\begin{remark}
	By setting $\tau = 0$ in \eqref{eq:tau_relation}, i.e., both error events are equally penalized, we obtain
     \begin{align*}
         E_\FA = D(p_{u^*} \| p_0 | p_X) = D(p_{u^*} \| p_1 | p_X) = E_\MD. 
     \end{align*}
    Moreover, from \eqref{eq:KLD_p0} at $\tau=0$,
    \begin{align*}
        D(p_{u^*} \| p_0 | p_X) = D(p_{u^*} \| p_1 | p_X) = -\kappa_0(u^*|p_X).
    \end{align*}
    That is, the error exponents are indeed $-\kappa_0(u^*|p_X)$, the $p_X$-weighted sum of Chernoff divergences. This special case at $\tau=0$ is the result in \cite[Thm.~1]{Chang--Wang--Erdogan--Bloch--2023} when the state space is binary. Further specializing it for the multiplicative state recovers \cite[Thm.~1]{Joudeh--Willems2022}.
\end{remark} 

\subsection{Converse proof}
Suppose that a rate-exponent tuple $(R, E_\FA, E_\MD)$ is achievable. That is, for any $\eps > 0$, there exists a length-$n$ codebook $\Cc^{(n)}$ such that $\frac{\log |\Cc^{(n)}|}{n} \ge R-\eps$, $P_e^{(n)} \le \epsilon$, and for every codeword $x^n \in \Cc^{(n)}$,
\begin{align*}
	\P( \Sh(Y^n, m) = 1 | M=m, S=0 ) &\le e^{-n(E_\FA - \eps)}, \\
	\P( \Sh(Y^n, m) = 0 | M=m, S=1 ) &\le e^{-n(E_\MD - \eps)}.
\end{align*}

Note that there are exponentially many codewords in $\Cc^{(n)}$, while the number of types of length $n$ is at most polynomial in $n$ \cite{Csiszar--Korner2011}. It implies that one can take a nonempty set of types $\Tc$ such that for all $p_X \in \Tc$, a subcodebook $\Cc_{p_X}^{(n)} := \{ x^n(m): \pi(x^n(m)) = p_X \} \subset \Cc^{(n)}$ satisfies
\begin{align*}
	\frac{ \log |\Cc_{p_X}^{(n)}| }{n} \ge  R - \eps - \delta_n
\end{align*}
for some $\delta_n > 0$ that tends to zero as $n\to\infty$ where $\pi(x^n)$ is the type of $x^n$.

In the sequel, we restrict our attention to $\Cc_{p_X}^{(n)}$. Let $M'$ be a message taking values uniformly random from the codewords in $\Cc_{p_X}^{(n)}$, and define a chosen codeword by $\Xt^n(m)$. Then, $\Xt^n(m) \sim \tilde{p}^n(\xt^n)$ and $\Xt_i\sim \tilde{p}(\xt)$ where
\begin{align*}
	\tilde{p}^n(x^n) &= \frac{\mathbf{1}\{x^n \in \Cc_{p_X}^{(n)} \} } {|\Cc_{p_X}^{(n)}|}, \\
	\tilde{p}_i(x) &= \frac{1}{|\Cc_{p_X}^{(n)}|} \sum_{x^n \in \Cc_{p_X}^{(n)}} \mathbf{1}\{x_i = x\}.
\end{align*}
That is, $\tilde{p}^n(\xt^n)$ is the uniform distribution over $\Cc_{p_X}^{(n)}$, and $\tilde{p}_i(\xt_i)$ is the $i$-th marginal distribution of symbols on $\Cc_{p_X}$. 
Also, let $\bar{p}(x)$ be the empirical distribution over the entire codebook $\Cc_{p_X}^{(n)}$ given by
\begin{align*}
	\bar{p}(x) &:= \frac{1}{n} \sum_{i=1}^n \tilde{p}_i(x) = p_X(x),
\end{align*}
where the last equality holds since all codewords in $\Cc_{p_X}^{(n)}$ are of type $p_X$. 

By applying Fano's inequality and standard converse steps, e.g.,~\cite[Chap.~7.2]{El-Gamal--Kim2011}, to the subcodebook $\Cc_{p_X}^{(n)}$, for every $s$,
\begin{align*}
	n(R-\eps - \delta_n) &\le \log |\Cc_{p_X}^{(n)}| \\
    &\le I(M'; \Yt^n) + n \eps_n \\
	&\le n I( \bar{p}_X, p_{\Yt|X,s}) + n\eps_n \\
    &= n I( p_X, p_{\Yt|X,s}) + n\eps_n.
\end{align*}
Both bounds for $s=0,1$ should hold, which gives the condition $R \le \min_{s=0,1} I(p_X, p_{\Yt|X,s}) + \eps + \delta_n + \eps_n.$

Next, we prove the bounds on the error exponents. Again, we first restrict our attention to the subcodebook $\Cc_{p_X}^{(n)}$. For binary hypothesis testing problems, the log-likelihood ratio test based on all collected information achieves the optimal tradeoff points of the error probabilities $P_\FA$ and $P_\MD$ \cite{Poor1994, MoulinV2019}. Moreover, by controlling the threshold parameter $\tau$, we can choose the operating point on the optimal tradeoff curve.

Define a binary decision random variable $T_s$, $s\in\{0,1\}$, 
\begin{align*}
	T_s = \begin{cases}
		0 & \text{if } \frac{1}{n} \LLR(Y^n | x^n) < \tau, \\
		1 & \text{if } \frac{1}{n} \LLR(Y^n | x^n) \ge \tau,
	\end{cases}
\end{align*}
where $\tau$ is the threshold that achieves the error exponents $E_\FA$, $E_\MD$. The subscript index $s$ indicates that the observation $Y^n|x^n$ is distributed as
\begin{align*}
	Y^n|x^n \sim \begin{cases}
		p_0(y^n|x^n) \text{ for } T_0,\\
		p_1(y^n|x^n) \text{ for } T_1.
	\end{cases}
\end{align*}
Also, let $u$ be such that
\begin{align}\label{eq:ustar}
	\mu_u := \E_{u}\left[ \frac{1}{n} \LLR(Y^n|x^n) \right] = \tau,
\end{align}
and similarly for some $\delta > 0$, let $v$ be such that 
\begin{align*}
	\mu_v = \E_{v}\left[ \frac{1}{n} \LLR(Y^n|x^n) \right] = \tau + \delta,
\end{align*}
where the underlying distribution $p_v$ is the mixture distribution similarly defined in \eqref{eq:geo_mix0}. Then, $T_v$ is also similarly defined with respect to observations following $p_v(y^n|x^n)$.

With the above definitions, 
\begin{align*}
	&D( p_v(Y^n|x^n) \| p_0(Y^n|x^n) )  \\  
	&\stackrel{(a)}{\ge} D( p(T_v|x^n)\| p(T_0|x^n) ) \\
	&= \P(T_v=0|x^n) \log \frac{ \P(T_v=0|x^n) }{ \P(T_0=0|x^n)} \\
	&\qquad \qquad + \P(T_v=1|x^n) \log \frac{ \P(T_v=1|x^n) }{ \P(T_0=1|x^n)} \\
	&= -H_2(T_v|x^n) - \P(T_v=0|x^n) \log \P(T_0=0|x^n) \\
	&\qquad \qquad - \P(T_v=1|x^n) \log \P(T_0=1|x^n) \\
	&\stackrel{(b)}{\ge} -1 - \P(T_v=1|x^n) \log \P(T_0=1|x^n),
\end{align*}
where $(a)$ follows from the data processing inequality since the binary random variable $T_s$ is a function of observations, and $(b)$ follows since the binary entropy function is upper bounded by $1$ and $\P(T_v=0|x^n) \log \P(T_0=0|x^n)< 0$. Rearranging terms and using the fact that $\P(T_v=1|x^n) \ge 1 - \frac{C}{n \delta^2}$ as shown in App.~\ref{app:appA}, we have
\begin{align*}
	&\log \P(T_0=1|x^n) \\
    &\ge \frac{-1 - D( p_v(Y^n|x^n) \| p_0(Y^n|x^n) )}{\P(T_v=1|x^n)} \\
	&\ge \frac{-1 - D( p_v(Y^n|x^n) \| p_0(Y^n|x^n) )}{ 1 - \frac{C}{n \delta^2} } \\
	&= (-1 - D( p_v(Y^n|x^n) \| p_0(Y^n|x^n) ) ) (1 + o(1)).
\end{align*}
Exponentiating both sides,
\begin{align*}
	&P_\FA(x^n) = \P(T_0=1|x^n) \\
	&\ge \exp ( - (1+D( p_v(Y^n|x^n) \| p_0(Y^n|x^n)))(1+o(1)) ).
\end{align*}
Since the additivity of the KL divergence for product distributions gives us 
\begin{align*}
	&D \left( p_v(Y^n|x^n) \| p_0(Y^n|x^n) \right) \\
	&= D \left( \prod_{i=1}^n p_u(y_i|x_i) \Bigg\| \prod_{i=1}^n p_0(y_i|x_i) \right) \\
	&= \sum_{i=1}^n D( p_u(Y_i|x_i) \| p_0(Y_i|x_i) ) \\
	&= n D( p_v(Y|X) \| p_0(Y|X) | p_X ),
\end{align*}
the above implies that for any $x^n \in \Cc_{p_X}^{(n)}$,
\begin{align*}
	P_\FA(x^n) \ge \exp\left( - n (D( p_v(Y|X) \| p_0(Y|X) | p_X ) + o(1)) \right).
\end{align*}
Finally, from two facts that $p_v \to p_u$ as $\delta \to 0$ and that $D( p_v(Y|X) \| p_0(Y|X) | p_X )$ is continuous in $p_v$, 
\begin{align*}
	P_\FA(x^n) \ge \exp\left( - n (D( p_u(Y|X) \| p_0(Y|X) | p_X ) + o(1)) \right)
\end{align*}
holds as $\delta \to 0$.

To obtain the bound on $P_\MD$, take $v$ such that $\mu_v = \tau - \delta$. By repeating the same steps for $D(p_v(Y^n|x^n) \| p_1(Y^n|x^n) )$ with $\delta \to 0$, we have
\begin{align*}
	P_\MD(x^n) \ge \exp\left( - n (D( p_u(Y|X) \| p_1(Y|X) | p_X ) + o(1)) \right).
\end{align*}
Since the above bounds hold for any $x^n \in \Cc_{p_X}^{(n)}$, it also holds that
\begin{align*}
	e^{-n (E_\FA-\epsilon)} &\ge \max_{ x^n \in \Cc^{(n)} } P_\FA(x^n) \\
	&\ge \max_{x^n\in \Cc_{p_X}^{(n)}} P_\FA(x^n) \\
	&\ge \exp\left( - n (D( p_u(Y|X) \| p_0(Y|X) | p_X ) + o(1)) \right),
\end{align*}
and
\begin{align*}
	e^{-n (E_\MD-\epsilon)} &\ge \max_{ x^n \in \Cc^{(n)} } P_\MD(x^n) \\
	&\ge \max_{x^n\in \Cc_{p_X}^{(n)}} P_\MD(x^n) \\
	&\ge \exp\left( - n (D( p_u(Y|X) \| p_1(Y|X) | p_X ) + o(1)) \right).
\end{align*}
Finally, we have the corresponding bounds on the exponents
\begin{align*}
	E_\FA \le D( p_u(Y|X) \| p_0(Y|X) | p_X ), \\
	E_\MD \le D( p_u(Y|X) \| p_1(Y|X) | p_X ).
\end{align*}

\subsection{Numerical Examples} \label{subsec:fixed_state_numerical_ex}
\textbf{Binary channel:} We begin with an example of a binary multiplicative state described in \cite{Joudeh--Willems2022}. This can be considered as an abstract model for the case that, if an obstacle is present, it completely disrupts the sensing path causing the detector to receive only noise. Specifically, we consider a state-independent communication channel and a state-dependent sensing channel such that
\begin{align*}
	\Yt_i &= x_i \oplus \Zt_i, \\
	Y_i &= S \cdot x_i \oplus Z_i,
\end{align*}
where $\Zt_i \sim \text{Bern}(p)$, $Z_i \sim \text{Bern}(q)$, and $\oplus$ is the modulo-$2$ sum. From the model, it can be observed that if $X_i=0$, the detector only observes natural noise regardless of the channel state; therefore, transmitting $X_i=1$ is the only action that contributes to detection. However, for communication purposes, the achievable rate is maximized with uniform random binary signals, deviating further from the binary symmetric channel capacity as the composition is biased towards input $X=1$. As a result, the tradeoff between data rate and sensing performance is determined by the symbol composition of the codewords, $p_X$. Then, for a given code composition, the  tradeoff between false alarm and missed detection error exponents is controlled by $p_u$, or equivalently, by the decision threshold $\tau = \tau(u)$.

\begin{figure}[!tb]
	\centering
	\resizebox{!}{19em}{\input{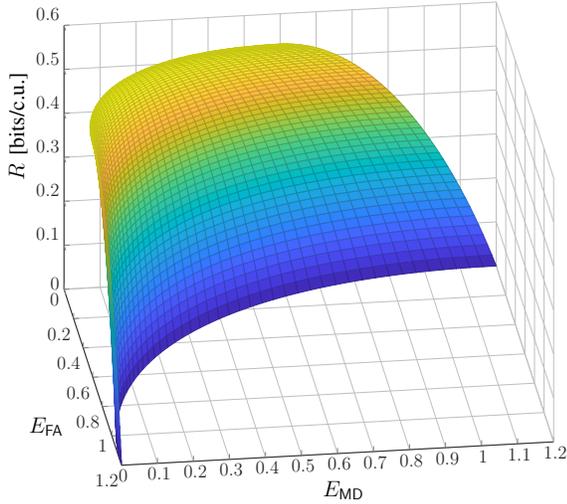}}
	\caption{Tradeoff boundary points of $\Rc$ for the binary multiplicative state channel with $p=0.1, q=0.2$. Units are bits.}
	\label{fig:binary_plot}
\end{figure}

In this setting, note that
\begin{align*}
	p_u(1|0) &= \frac{p_0^{1-u}(1|0) p_1^u(1|0)}{p_0^{1-u}(0|0) p_1^u(0|0) + p_0^{1-u}(1|0) p_1^u(1|0)} \\
	&= \frac{ q^{1-u} q^u  }{ (1-q)^{1-u} (1-q)^{u} + q^{1-u} q^u } \\
	&= q, \\
	p_u(1|1) &= \frac{p_0^{1-u}(1|1) p_1^u(1|1)}{p_0^{1-u}(0|1) p_1^u(0|1) + p_0^{1-u}(1|1) p_1^u(1|1)} \\
	&= \frac{ q^{1-u} (1-q)^u  }{ (1-q)^{1-u} q^{u} + q^{1-u} (1-q)^u }.
\end{align*}
Then, letting $t := p_X(1)$ for brevity and applying Thm.~\ref{thm:theorem1},
\begin{align*}
	R &\le I(X, \Yt) \\
	&= H_2(t * p) - H_2(p), \\
	E_\FA &\le D( p_u(Y|X) \| p_1(Y|X) | p_X ) \\
	&= (1-t) D(p_u(Y|0) \| p_1(Y|0)) + t D(p_u(Y|1) \| p_1(Y|1)) \\
	&= t D_2\left(  \frac{ q^{1-u} (1-q)^u  }{ (1-q)^{1-u} q^{u} + q^{1-u} (1-q)^u }  \bigg\| 1-q \right), \\
	E_\MD &\le D( p_u(Y|X) \| p_0(Y|X) | p_X ) \\
	&= (1-t) D(p_u(Y|0) \| p_0(Y|0)) + \alpha D(p_u(Y|1) \| p_0(Y|1)) \\
	&= t D_2 \left( \frac{ q^{1-u} (1-q)^u  }{ (1-q)^{1-u} q^{u} + q^{1-u} (1-q)^u }  \bigg\|  q \right),
\end{align*}
where $t * p = t(1-p) + (1-t)p$ and $D_2(a \| b)$ is the KL divergence between two binary distributions. The region is plotted in Fig.~\ref{fig:binary_plot}.

As stated, for the special case with $u=0.5$, i.e., both error events are equally penalized, it recovers the rate region given in~\cite{Joudeh--Willems2022} as follows.
\begin{corollary}[{\cite[Thm.~1]{Joudeh--Willems2022}}] \label{cor:willems_example}
	For the binary multiplicative channel with equal exponent constraint, $E := E_{\FA}=E_{\MD}$, the optimal trade-off region is given by the set of $(R,E)$ pairs such that
	\begin{align*}
		R &\le H_2(t * p)-H_2(p),\\
		E &\le t D_2(0.5\| q).
	\end{align*}
    for some $t \in [0.5, 1]$, where $t * p = t(1-p) + (1-t)p$.
\end{corollary}

\textbf{Gaussian channel:} Next, consider an AWGN communication channel
\begin{align*}
	\Yt_i &= x_i + \Zt_i,
\end{align*}
and an AWGN sensing channel
\begin{align*}
	Y_i &= x_i + Z_i, \text{ if } S=0,\\
	Y_i &= hx_i + Z_i, \text{ if } S=1,
\end{align*}
where we assume an average input power constraint $\sum_{i=1}^n x_i^2\le nP$, $\Zt_i \sim \mathcal{N}(0,1)$, $Z_i \sim \mathcal{N}(0,1)$, and $h \ne 1$. This is also the case when $p_{\Yt|X,S=0} = p_{\Yt|X,S=1} =:  p_{\Yt|X}$ and can be considered as an abstract model for obstacle detection in which an obstacle alters the detector's wireless propagation path. In this model, the unknown state is amplified by the transmitted signal. Therefore, transmitting $X_i$ as large as possible within the power constraint is optimal for sensing. Moreover, as shown in Thm.~\ref{thm:theorem1}, error exponents are conditional KL divergences, which will be shown to be proportional to $x^2$ given $x$. This implies that the error exponents are proportional to $\E[X^2]$, i.e., the average transmission power. Hence, by using a Gaussian codebook with average power $P$, one can achieve the optimal channel capacity as well as the optimal detection error exponents. For a fixed input distribution, the tradeoff between the false alarm and missed detection probabilities are further determined by $p_u$ or equivalently by $\tau$.

Assuming $X \sim \mathcal{N}(0, P)$, note that
\begin{align*}
	R \le \frac{1}{2} \log (1 + P).
\end{align*}
To evaluate the exponents, first note that
\begin{align*}
	p_u(y|x) &= \frac{ \frac{1}{\sqrt{2\pi}} \exp\left( -\frac{(1-u)(y-x)^2}{2} \right)  \exp\left( -\frac{u(y-hx)^2}{2} \right) }{ \int_{\mathbb{R}} \frac{1}{\sqrt{2\pi}} \exp\left( -\frac{(1-u)(y-x)^2}{2} \right)  \exp\left( -\frac{u(y-hx)^2}{2} \right) dy } \\
	&= \frac{1}{\sqrt{2\pi}} \exp\left( -\frac{ (y - (1-u+hu)x)^2}{2} \right) \\
	&= \Nc((1-u+hu)x, 1).
\end{align*}
With the mixture distribution at hand, we have
\begin{align}
	D&(p_u(y|X=x) \| p_1(y|X=x)) \nn \\
	&= D(\Nc((1-u+hu)x, 1) \| \Nc(hx, 1)) \nn \\
	&= \frac{(1-u)^2 (1-h)^2 x^2}{2}, \label{eq:exp_x_1} \\
	D&(p_u(y|X=x) \| p_0(y|X=x)) \nn \\
	&= D(\Nc((1-u+hu)x, 1) \| \Nc(x, 1)) \nn \\
	&= \frac{u^2 (1-h)^2 x^2}{2}. \label{eq:exp_x_2}
\end{align}
Then, we can evaluate Thm.~\ref{thm:theorem1} under the Gaussian input distribution $X \sim \mathcal{N}(0, P)$ which gives the following region:
\begin{align}
	R &\le \frac{1}{2} \log (1 + P), \nn\\
	E_\FA &\le D( p_u(Y|X) \| p_1(Y|X) | p_X ) = \frac{(1-u)^2 (1-h)^2 P}{2}, \nn\\
	E_\MD &\le D( p_u(Y|X) \| p_0(Y|X) | p_X ) = \frac{u^2 (1-h)^2 P}{2}. \label{eq:region_gaussian}
\end{align}
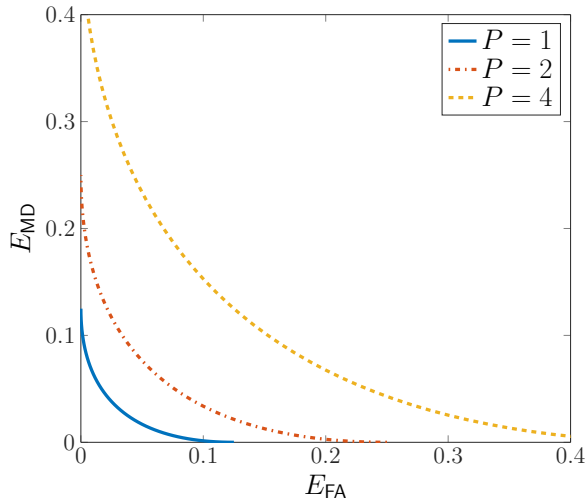
\begin{figure}[!t]
	\centering
	\resizebox{!}{19em}{
%
%
%
\definecolor{mycolor1}{rgb}{0.00000,0.44700,0.74100}%
\definecolor{mycolor2}{rgb}{0.85000,0.32500,0.09800}%
\definecolor{mycolor3}{rgb}{0.92900,0.69400,0.12500}%
\begin{tikzpicture}

\begin{axis}[%
width=4.13819444444445in,
height=3.61666666666667in,
scale only axis,
separate axis lines,
every outer x axis line/.append style={white!15!black},
every x tick label/.append style={font=\Large\color{white!15!black}},
xmin=0,
xmax=0.4,
xtick distance=0.1,
xlabel={\LARGE$E_{\FA}$},
every outer y axis line/.append style={white!15!black},
every y tick label/.append style={font=\Large\color{white!15!black}},
ymin=0,
ymax=0.4,
ytick distance=0.1,
ylabel={\LARGE$E_{\MD}$},
legend style={draw=white!15!black,fill=white,legend cell align=left}
]
\addplot [color=mycolor1,solid,line width=2.0pt]
  table[row sep=crcr]{%
0.125	6.16297582203915e-33\\
0.119950020824656	5.20616409829249e-05\\
0.115004164931279	0.000208246563931697\\
0.110162432319867	0.000468554768846317\\
0.105424822990421	0.000832986255726784\\
0.10079133694294	0.0013015410245731\\
0.096261974177426	0.00187421907538526\\
0.0918367346938775	0.00255102040816327\\
0.0875156184922948	0.00333194502290713\\
0.083298625572678	0.00421699291961683\\
0.0791857559350271	0.00520616409829239\\
0.0751770095793419	0.00629945855893378\\
0.0712723865056226	0.00749687630154103\\
0.0674718867138692	0.00879841732611413\\
0.0637755102040816	0.0102040816326531\\
0.0601832569762599	0.0117138692211579\\
0.056695127030404	0.0133277800916285\\
0.0533111203665139	0.015045814244065\\
0.0500312369845897	0.0168679716784673\\
0.0468554768846314	0.0187942523948355\\
0.0437838400666389	0.0208246563931695\\
0.0408163265306122	0.0229591836734694\\
0.0379529362765514	0.0251978342357351\\
0.0351936693044565	0.0275406080799667\\
0.0325385256143274	0.0299875052061641\\
0.0299875052061641	0.0325385256143274\\
0.0275406080799667	0.0351936693044565\\
0.0251978342357351	0.0379529362765514\\
0.0229591836734694	0.0408163265306122\\
0.0208246563931695	0.0437838400666389\\
0.0187942523948355	0.0468554768846314\\
0.0168679716784673	0.0500312369845897\\
0.015045814244065	0.0533111203665139\\
0.0133277800916285	0.056695127030404\\
0.0117138692211579	0.0601832569762599\\
0.0102040816326531	0.0637755102040816\\
0.00879841732611412	0.0674718867138692\\
0.00749687630154103	0.0712723865056226\\
0.00629945855893378	0.0751770095793419\\
0.00520616409829238	0.0791857559350271\\
0.00421699291961683	0.083298625572678\\
0.00333194502290713	0.0875156184922948\\
0.00255102040816327	0.0918367346938775\\
0.00187421907538526	0.096261974177426\\
0.0013015410245731	0.10079133694294\\
0.000832986255726785	0.105424822990421\\
0.000468554768846317	0.110162432319867\\
0.000208246563931698	0.115004164931279\\
5.2061640982925e-05	0.119950020824656\\
6.16297582203915e-33	0.125\\
};
\addlegendentry{\LARGE$P=1$};

\addplot [color=mycolor2,dash pattern=on 1pt off 3pt on 3pt off 3pt,line width=2.0pt]
  table[row sep=crcr]{%
0.25	1.23259516440783e-32\\
0.239900041649313	0.00010412328196585\\
0.230008329862557	0.000416493127863394\\
0.220324864639733	0.000937109537692634\\
0.210849645980841	0.00166597251145357\\
0.201582673885881	0.0026030820491462\\
0.192523948354852	0.00374843815077052\\
0.183673469387755	0.00510204081632654\\
0.17503123698459	0.00666389004581426\\
0.166597251145356	0.00843398583923367\\
0.158371511870054	0.0104123281965848\\
0.150354019158684	0.0125989171178676\\
0.142544773011245	0.0149937526030821\\
0.134943773427738	0.0175968346522283\\
0.127551020408163	0.0204081632653061\\
0.12036651395252	0.0234277384423157\\
0.113390254060808	0.026655560183257\\
0.106622240733028	0.03009162848813\\
0.100062473969179	0.0337359433569346\\
0.0937109537692628	0.037588504789671\\
0.0875676801332778	0.041649312786339\\
0.0816326530612245	0.0459183673469388\\
0.0759058725531029	0.0503956684714702\\
0.0703873386089129	0.0550812161599334\\
0.0650770512286547	0.0599750104123282\\
0.0599750104123282	0.0650770512286547\\
0.0550812161599334	0.070387338608913\\
0.0503956684714702	0.0759058725531029\\
0.0459183673469388	0.0816326530612245\\
0.0416493127863391	0.0875676801332778\\
0.037588504789671	0.0937109537692628\\
0.0337359433569346	0.100062473969179\\
0.03009162848813	0.106622240733028\\
0.026655560183257	0.113390254060808\\
0.0234277384423157	0.12036651395252\\
0.0204081632653061	0.127551020408163\\
0.0175968346522282	0.134943773427738\\
0.0149937526030821	0.142544773011245\\
0.0125989171178676	0.150354019158684\\
0.0104123281965848	0.158371511870054\\
0.00843398583923366	0.166597251145356\\
0.00666389004581426	0.17503123698459\\
0.00510204081632654	0.183673469387755\\
0.00374843815077053	0.192523948354852\\
0.0026030820491462	0.201582673885881\\
0.00166597251145357	0.210849645980841\\
0.000937109537692634	0.220324864639733\\
0.000416493127863396	0.230008329862557\\
0.00010412328196585	0.239900041649313\\
1.23259516440783e-32	0.25\\
};
\addlegendentry{\LARGE$P=2$};

\addplot [color=mycolor3,dashed,line width=2.0pt]
  table[row sep=crcr]{%
0.5	2.46519032881566e-32\\
0.479800083298625	0.000208246563931699\\
0.460016659725114	0.000832986255726789\\
0.440649729279467	0.00187421907538527\\
0.421699291961682	0.00333194502290714\\
0.403165347771762	0.0052061640982924\\
0.385047896709704	0.00749687630154104\\
0.36734693877551	0.0102040816326531\\
0.350062473969179	0.0133277800916285\\
0.333194502290712	0.0168679716784673\\
0.316743023740108	0.0208246563931695\\
0.300708038317368	0.0251978342357351\\
0.285089546022491	0.0299875052061641\\
0.269887546855477	0.0351936693044565\\
0.255102040816326	0.0408163265306123\\
0.240733027905039	0.0468554768846314\\
0.226780508121616	0.053311120366514\\
0.213244481466056	0.0601832569762599\\
0.200124947938359	0.0674718867138693\\
0.187421907538526	0.075177009579342\\
0.175135360266556	0.0832986255726781\\
0.163265306122449	0.0918367346938776\\
0.151811745106206	0.10079133694294\\
0.140774677217826	0.110162432319867\\
0.130154102457309	0.119950020824656\\
0.119950020824656	0.130154102457309\\
0.110162432319867	0.140774677217826\\
0.10079133694294	0.151811745106206\\
0.0918367346938776	0.163265306122449\\
0.0832986255726781	0.175135360266556\\
0.075177009579342	0.187421907538526\\
0.0674718867138693	0.200124947938359\\
0.0601832569762599	0.213244481466056\\
0.053311120366514	0.226780508121616\\
0.0468554768846314	0.240733027905039\\
0.0408163265306123	0.255102040816326\\
0.0351936693044565	0.269887546855477\\
0.0299875052061641	0.285089546022491\\
0.0251978342357351	0.300708038317368\\
0.0208246563931695	0.316743023740108\\
0.0168679716784673	0.333194502290712\\
0.0133277800916285	0.350062473969179\\
0.0102040816326531	0.36734693877551\\
0.00749687630154105	0.385047896709704\\
0.0052061640982924	0.403165347771762\\
0.00333194502290714	0.421699291961682\\
0.00187421907538527	0.440649729279467\\
0.000832986255726791	0.460016659725114\\
0.0002082465639317	0.479800083298625\\
2.46519032881566e-32	0.5\\
};
\addlegendentry{\LARGE$P=4$};

\end{axis}
\end{tikzpicture}%
}
	\caption{Projection of $\Rc$ on the $E_\MD$-$E_\FA$ plane for the Gaussian example with $P=1, P=2$ and $P=4$. We set $h=0.5$.}
	\label{fig:gaussian_plot}
\end{figure}

We note that the region~\eqref{eq:region_gaussian} is in fact $\Rc$, i.e., the Gaussian distribution achieves the largest region over all possible input distributions. To see this, note that for each $x$, \eqref{eq:exp_x_1} and \eqref{eq:exp_x_2} hold. That means, for any distribution satisfying the constraint $\E(X^2)\le P$, the exponents are upper bounded by
\begin{align*}
	D(p_u(Y|X) \| p_1(Y|X)|p_X) &\le \frac{(1-u)^2 (1-h)^2 P}{2}, \\
	D(p_u(Y|X) \| p_0(Y|X)|p_X) &\le \frac{u^2 (1-h)^2 P}{2}.
\end{align*}
Since Gaussian distribution achieves the equalities and maximizes the mutual information, it attains $\Rc$ in Thm.~\ref{thm:theorem1}. The region $\Rc$ is plotted in Fig.~\ref{fig:gaussian_plot} for a few selected values of $P$.

\textbf{Vector Gaussian channel:} Next, consider an AWGN MIMO communication channel
\begin{align*}
	\tilde{\Yv}_i &= \tilde H \xv_i + \tilde{\Zv}_i,
\end{align*}
and an AWGN MIMO sensing channel
\begin{align*}
	\Yv_i &= H_0 \xv_i + \Zv_i, \text{ if } S=0,\\
	\Yv_i &= H_1 \xv_i + \Zv_i, \text{ if } S=1,
\end{align*}
where we assume an average input power constraint $\sum_{i=1}^n \|x_i\|^2\le nP$, and independent AWGN $\Zt_i \sim \mathcal{N}(0,I)$, $Z_i \sim \mathcal{N}(0,I)$. For the vector Gaussian channel, we have
\begin{align*}
    p_u(\yv|\xv) = \Nc((1-u)H_0\xv+uH_1\xv, I).
\end{align*}
Then, by choosing a Gaussian input with $X\sim \Nc(0, \Sigma_X)$, where $\tr(\Sigma_X)\le P$, we have the following corollary.
\begin{corollary}\label{cor:vector_gaussian}
For the vector Gaussian channel, a rate-exponent tuple $(R,E_{\FA}, E_{\MD})$ is achievable  if
\begin{align*}
    R &< \frac{1}{2}\log\left|I+\tilde{H}\Sigma_X\tilde{H}^T\right|\\
    E_{\FA} &< \frac{(1-u)^2}{2} \tr(\Gamma \Sigma_X ),\\
    E_{\MD} &< \frac{u^2}{2} \tr(\Gamma \Sigma_X ),
\end{align*}
for some covariance matrix $\Sigma_X$ such that $\tr(\Sigma_X)\le P$ and $u\in(0,1)$, where $\Gamma = (H_1-H_0)^T(H_1-H_0)$.    
\end{corollary}

Some remarks on the region in Cor.~\ref{cor:vector_gaussian} are as follows. We begin by identifying the optimal extreme points in the region, i.e., $E_\FA=E_\MD = 0$ and $R=0$ cases. For $E_\FA=E_\MD = 0$ case, the maximum rate is the standard vector Gaussian channel capacity which is attained by beamforming with respect to the SVD of $\tilde H$ and waterfilling over the singular values~\cite{Telatar1999}. For the other corner point case $R=0$, the optimal detection error exponents are attained by allocating all the power to the maximum singular value of $(H_1-H_0)$ and beamforming in the direction of the corresponding unitary vector. Overall, the input covariance matrix governs the tradeoff between rate and error exponents (but not between the error exponents). Thus, there is an intrinsic multiplexing--diversity tradeoff, where allocating power via waterfilling on multiple sub-channels is optimal for communication, while investing all power and beam direction on the best sub-channel of the difference matrix $(H_1-H_0)$ is optimal for detection error. 
Furthermore, for a fixed covariance $\Sigma_X$, we can 
find that the tradeoff between the error exponents $E_\FA$ and $E_\MD$ is completely characterized by $u\in(0,1)$.

\begin{figure}[!tb]
	\centering
      \psfrag{md}{$E_\MD$}
     \psfrag{fa}{$E_\FA$}
    \psfrag{ra}{$R$ [bits/c.u.]}
       \includegraphics[width=20em]{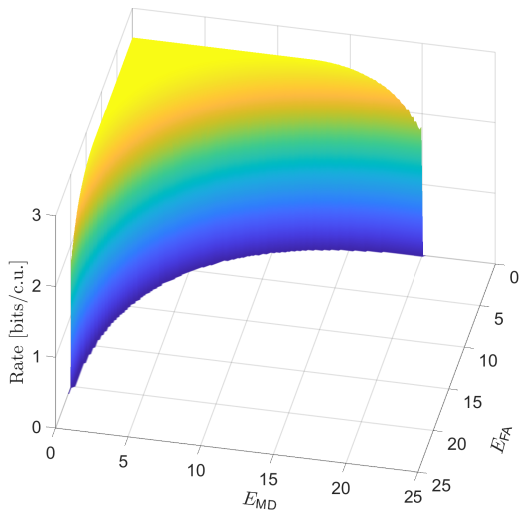}
	\caption{Achievable points $(R, E_{\FA}, E_{\MD})$ for the vector Gaussian channel with $P=10$ and channel gains given in~\eqref{eq:mimo_channel_gains}.}
	\label{fig:mimo_ex}
\end{figure}

We plot an example case for the vector setup in Fig.~\eqref{fig:mimo_ex} where we assume power constraint $P=10$ and
\begin{align}
    (H_1-H_0) = \begin{bmatrix}
        2 & 0 \\
        0 & 1 
    \end{bmatrix},\quad  \tilde{H} = \frac{1}{\sqrt{2}}\begin{bmatrix}
        1 & 1 \\
        1 & -1 
    \end{bmatrix}.\label{eq:mimo_channel_gains}
\end{align}

\section{Tradeoff for i.i.d.~States} \label{sec:iid_state}
Consider a scenario where the state changes every symbol in an i.i.d.~manner, and there are separate error probability constraints for probabilities of false alarm and missed detection, respectively. In this case, unlike in Sec.~\ref{sec:main results}, each state must be detected based on a single observation. Note that if $\Yc$ is finite, then the likelihood ratio for a single observation, $\frac{p_1(y|x)}{p_0(y|x)}$, also takes a finite number of values. Consequently, using deterministic detection rules results in only a finite number of probability pairs of false alarm and missed detection. However, we wish to control these error probabilities continuously. To this end, we allow the detector to perform randomized detection, i.e., for a given $y$, use a detection rule such that 
\begin{align*}
	\Sh_x(y) = \begin{cases}
		\sh_1(y|x) & \text{with probability } \eta, \\
		\sh_2(y|x) & \text{with probability } 1-\eta,
	\end{cases}
\end{align*}
where $\hat{s}_1, \hat{s}_2$ are deterministic detection rules. Or equivalently, the randomized detection rule $\Sh_x$ can be thought of as a probability mapping on a test channel $p_{\Sh|Y,X}$. By varying $\sh_1, \sh_2, \eta$, we can control $(P_\FA, P_\MD)$ for $\hat{S}$ in a continuous manner.\footnote{This randomization changes only a constant factor in $P_\FA, P_\MD$, which implies that the error exponents remain unchanged. Hence, further considering randomization detection rules in Sec.~\ref{sec:main results} does not change error exponents.}

Also, with a slight abuse of notation, let $P_{\FA,x}(\Sh), P_{\D,x}(\Sh)$ respectively be the probabilities of false alarm and correct detection by using detection rule $\Sh$, i.e.,
\begin{align*}
	P_{\FA,x}(\Sh) &= \P[ \Sh = 1 | X=x, S=0 ], \\
	P_{\D,x}(\Sh) &= \P[ \Sh = 1 | X=x, S=1 ].
\end{align*}
This is well defined since the channel is memoryless and $Y$ is independent of others if $X=x, S=s$ are given. Note that as the state varies i.i.d.~according to $p(s)$, the marginalized channel $p(\yt|x)$ can be computed as follows.
\begin{align*}
	p(\yt|x) = \sum_{s} p(\yt,s|x) = \sum_{s} p(s)p(\yt|x,s).
\end{align*}
Then, the rate-probability region $\Rc$ is implicitly characterized as follows. The proof of the next theorem is given in the next subsection. From the proof, the following corollary is immediate.
\begin{theorem} \label{thm:iid_state}
	The rate-probability region $\Rc$ is the set of tuples $(R, \alpha, \beta)$ such that for some $p_X, p_{\Sh|Y,X}$,
	\begin{align}
		R &\le I(p_X, p_{\Yt|X}), \nn \\
		\beta &\le \sum_{x} p_X(x) P_{\D,x}(\Sh_x), \label{eq:beta_ineq}
	\end{align}
	where $\{\hat{S}_x\}_{x\in\Xc}$ are subject to
	\begin{align}
		\sum_{x} p_X(x) P_{\FA,x}(\Sh_x) \le \alpha. \label{eq:alpha_ineq}
	\end{align}
\end{theorem}
\begin{corollary}
A randomized detector is necessary in general to achieve the boundary points of $\Rc$.
\end{corollary}

Since this is an implicit characterization, it is important to address how to compute the region explicitly. Note that for each $x$, the tradeoff between $P_{\D,x}$ and $P_{\FA,x}$ is represented by the receiver operating characteristic (ROC) curve. The curve is fully determined by two likelihoods $p_0(y|x)=p(y|x,s=0), p_1(y|x)=p(y|x,s=1)$, and $P_{\D,x}$ is concave in $P_{\FA,x}$ 
\cite{MoulinV2019}. Given these properties, for a fixed $p_X$, \eqref{eq:beta_ineq} and \eqref{eq:alpha_ineq} are a convex program with variables $\{P_{\D,x}, P_{\FA,x}\}_{x \in \Xc}$ coupled together by $|\Xc|$ ROCs:
\begin{align*}
    \max &\sum_{x} p_X(x) P_{\D,x}(\Sh_x) \\
    \text{subject to } &\sum_{x} p_X(x) P_{\FA,x}(\Sh_x) \le \alpha.
\end{align*}
Therefore, for each $\alpha \in (0,1)$, we can numerically find the optimal $\{P_{\D,x}^*, P_{\FA,x}^*\}_{x \in \Xc}$, or equivalently, the optimal (possibly randomized) detection rule $\hat{S}_x(y)$ for each $x$. In the following, we demonstrate that this solution can be obtained by a waterfilling algorithm, e.g., \cite{BoydV2004, CoverT1991}. Formally, let $f$ be the objective function, i.e., 
\begin{align*}
	f &:= \sum_{x} p_X(x) P_{\D,x}(\Sh_x).
\end{align*}
Note that increasing $P_{\FA, x}$ by $\frac{\Delta}{p_X(x)}$ increases the entire probability of false alarm by $\Delta$. It increases the objective by 
\begin{align*}
	\frac{\Delta}{p_X(x)} \cdot \frac{\partial f}{\partial P_{\FA, x}} &= \frac{\Delta}{p_X(x)} \cdot \frac{p_X(x) \partial P_{\D, x}}{\partial P_{\FA, x}} \\
    &= \Delta \frac{\partial P_{\D, x}}{\partial P_{\FA, x}}.
\end{align*}
Hence, the optimal solution $\{P_{\D,x}^*, P_{\FA,x}^*\}_{x \in \Xc}$ can be found by incrementally pouring a small amount of probability of false alarm (e.g., water) into the bin that yields the greatest objective gain at each iteration. The algorithm for given $p_X$ and $\alpha$ is formally described in Algorithm \ref{alg:waterfilling}. At each iteration, the algorithm calculates the expected detection probability gain for each input symbol, considering the tradeoff within the false alarm budget, i.e., the slope of the ROC curve. It then allocates the false alarm budget to the input symbol that yields the largest detection probability gain.

\begin{algorithm}[t]
	\caption{Waterfilling algorithm for \eqref{eq:beta_ineq} and \eqref{eq:alpha_ineq}} \label{alg:waterfilling}
	\KwIn{$p_X$, $|\Xc|$ ROC curves, constraint $\alpha$, increment $\Delta$}
	\KwResult{Optimal $\{P_{\D,x}^*, P_{\FA,x}^*\}_{x \in \Xc}$ for \eqref{eq:beta_ineq} and \eqref{eq:alpha_ineq}}
	\textbf{Initialization: $P_{\FA,x} \leftarrow 0$ for all $x$ } \\
	\While{ $\sum_{x} p_X(x) P_{\FA,x}(\Sh_x) \le \alpha$ }{
		for every $x$, \\
		\eIf{$P_{\FA, x} = 1$}{
			$g(x) \leftarrow 0$ ~~~ \CommentSty{\footnotesize so that this $x$ is not chosen}
		}{
			$g(x) \leftarrow \frac{\partial P_{\D, x}}{\partial P_{\FA, x}}$
		}
		
		pick $x^\circ \leftarrow \argmax_{x} g(x)$
		$P_{\FA,x^\circ} \leftarrow P_{\FA,x^\circ} + \frac{\Delta}{p_X(x^\circ)}$
	}
	$P_{\FA,x}^* \leftarrow P_{\FA,x}$ for every $x$ \\
	find corresponding $P_{\D,x}^*$ from the $x$-th ROC curve
\end{algorithm}

This theorem can also be viewed from capacity-distortion perspective as in \cite{Ahmadipour--Kobayashi--Wigger--Caire2022}. By considering two distortion metrics $d_\FA(s, \sh)$ and $d_\MD(s, \sh) = 1-d_\D(s, \sh)$ such that
\begin{align*}
    d_\FA(s, \sh) &= \begin{cases}
        1 & \text{if } (s, \sh) = (0, 1), \\
        0 & \text{otherwise},
    \end{cases} \\
    d_\MD(s, \sh) &= \begin{cases}
        1 & \text{if } (s, \sh) = (1, 0), \\
        0 & \text{otherwise},
    \end{cases},
\end{align*}
the theorem can be recast as follows.
\begin{prop}[{Restatement of Thm.~\ref{thm:iid_state}}]
    The rate-probability region $\Rc$ is the set of tuples $(R, \alpha, \beta)$ such that for some $p_X, p_{\Sh|Y,X}$,
	\begin{align*}
		R &\le I(p_X, p_{\Yt|X}), \\
        \E[ d_\FA(S, \Sh(X, Y)) ] &\le \alpha, \\
		\E[ d_\MD(S, \Sh(X, Y)) ] &\le 1-\beta.
	\end{align*}
\end{prop}
Despite the equivalence, our detection-perspective characterization in Thm.~\ref{thm:iid_state} allows a more explicit understanding of the expression and optimization process.

It should be also remarked that randomized decision rules are necessary for our characterization, whereas deterministic rules suffice in \cite[Thm.~1]{Ahmadipour--Kobayashi--Wigger--Caire2022}. This difference stems from the objective. Recall that specializing in detection error probability, the goal in \cite{Ahmadipour--Kobayashi--Wigger--Caire2022} is to minimize the average error probability, the weighted sum of $P_{\FA}$ and $P_{\MD}$, corresponding to Bayesian detection. In this scenario, if the likelihood ratio exactly equals the decision threshold, the tie can be broken arbitrarily without affecting the average error probability. In contrast, our goal is to minimize $P_{\FA}$ and $P_{\MD}$ separately, as in Neyman-Pearson detection, where $P_{\FA}$ and $P_{\MD}$ are in general affected by a tie-breaking rule.

\subsection{Proof of Thm.~\ref{thm:iid_state}}
\textit{Achievability:} As in the achievability proof for Thm.~\ref{thm:theorem1}, we fix $p_X$ and use a constant composition code (CCC) with composition $p_X$. Since the effective channel seen by the receiver is
\begin{align*}
	p(\yt|x) &= \sum_{s} p(\yt,s|x) \\
    &= \sum_{s} p(s)p(\yt|x,s),
\end{align*}
i.e., an arbitrarily varying channel (AVC), the receiver can decode the message with an arbitrarily small error if
\begin{align*}
	R < I(p_X, p_{\Yt|X}),
\end{align*}
as block length $n \to \infty$. As it is a standard coding theorem, we refer to \cite{Csiszar--Korner2011} for details.

For state detection, the detector uses a symbolwise input-dependent detection rule $\hat{S}_x(y)$. This is feasible because the input signal is shared with the detector. Consequently, the $i$-th estimate depends only on $x_i$ and $y_i$, and therefore,
\begin{align*}
	P^{(n)}_{\FA,i}(x^n) &= \P[ \hat{S}_i(Y^n, x^n) = 1 | X^n = x^n, S_i = 0 ] \\
	&\stackrel{(a)}{=} \P[ \hat{S}_{x_i}(Y_i) = 1 | X^n = x^n, S_i = 0 ] \\
	&\stackrel{(b)}{=} \P[ \hat{S}_{x_i}(Y_i) = 1 | X_i = x_i, S_i = 0 ] \\
	&\stackrel{(c)}{=} \P[ \hat{S}_{x}(Y) = 1 | X = x, S = 0 ],
\end{align*}
where $(a)$ follows since our detection rule is symbowise and dependent on input, $(b)$ follows since if $X_i = x_i$ is given, then the probability of $Y_i$ is independent of $X_{-i}$ due to the channel being memoryless, and $(c)$ follows since the channel is i.i.d. and the estimator does not depend on the time index. Then, the time-averaged false alarm probability \eqref{eq:iid_FA_constraint} can be rewritten as follows:
\begin{align*}
	&\frac{1}{n} \sum_{i=1}^n P^{(n)}_{\FA,i}(x^n) \\
	&= \frac{1}{n} \sum_{i=1}^n \sum_{x \in \Xc} \mathbf{1}\{ X_i = x \} \P[ \hat{S}_{x}(Y) = 1 | X = x, S = 0 ] \\
	&\stackrel{(d)}{=} \sum_{x \in \Xc} p_X(x) \P[ \hat{S}_{x}(Y) = 1 | X = x, S = 0 ] \\
	&= \sum_{x \in \Xc} p_X(x) P_{\FA, x}(\hat{S}_x).
\end{align*}
where $(d)$ follows since every codeword has a constant composition of $p_X$. The probability of correct detection can be similarly rewritten as
\begin{align*}
	\frac{1}{n} \sum_{i=1}^n P^{(n)}_{\D,i}(x^n) = \sum_{x \in \Xc} p_X(x) P_{\D, x}(\hat{S}_x).
\end{align*}
Therefore, achieving \eqref{eq:iid_PD_constraint} and \eqref{eq:iid_FA_constraint} is indeed the same as \eqref{eq:beta_ineq} and \eqref{eq:alpha_ineq}.

\textit{Converse:} Suppose that for some $\alpha, \beta \in [0,1]$, a rate-probability tuple $(R, \alpha, \beta)$ is achievable by a codebook $\Cc^{(n)}$. That is, for any $\eps > 0$, there exists a codebook $\Cc^{(n)}$ of length-$n$ such that $\frac{\log |\Cc^{(n)}|}{n} \ge R-\eps$, $P_e^{(n)} \le \epsilon$, and for every $x^n \in \Cc^{(n)}$,
\begin{align}
	\frac{1}{n} \sum_{i=1}^n P^{(n)}_{\D,i}(x^n) \ge \beta-\epsilon, \label{eq:beta_constraint} \\
	\frac{1}{n} \sum_{i=1}^n P^{(n)}_{\FA,i}(x^n) \le \alpha+\epsilon. \label{eq:alpha_constraint}
\end{align}

Recalling the converse proof for Thm.~\ref{thm:theorem1}, note that there exists a subcode $\Cc_{p_X}^{(n)}$ of type $p_X$ that satisfies
\begin{align*}
	\frac{ \log |\Cc_{p_X}^{(n)}| }{n} \ge  R - \eps - \delta_n
\end{align*}
for some $\delta_n > 0$ that tends to zero as $n \to \infty$. Then, by applying standard converse steps~\cite{El-Gamal--Kim2011} to the subcode $\Cc_{p_X}^{(n)}$,
\begin{align*}
	n(R-\eps - \delta_n) &\le \log |\Cc_{p_X}^{(n)}| \\
	&\le n I(p_X, p_{\Yt|X}) + n\eps_n
\end{align*}
for some $\epsilon_n \to 0$ as $n \to \infty$, where
\begin{align*}
	p(\yt|x) = \sum_{s} p(\yt,s|x) = \sum_{s} p(s)p(\yt|x,s).
\end{align*}
In other words,
\begin{align*}
	R \le I(p_X, p_{\Yt|X}) + \eps + \delta_n + \eps_n.
\end{align*}

To detect $S_i$, it is optimal for the detector to perform a (possibly randomized) likelihood ratio test based on all information that the detector has. For brevity, let $\LR(y_i | x^n, m)$ be
\begin{align*}
	\LR(y_i | x^n, m) = \frac{ p(y_i | x^n, m, S_i=1) }{ p(y_i | x^n, m, S_i=0) }.
\end{align*}
Then, the optimal detection rule for the $i$-th state is
\begin{align*}
	\hat{S}_i = \begin{cases}
		1 & \text{if } \LR(y_i | x^n, m) > \tau_i, \\
		1 \text{ with probability } \eta_i & \text{if } \LR(y_i | x^n, m) = \tau_i, \\
		0 & \text{if } \LR(y_i | x^n, m) < \tau_i,
	\end{cases}
\end{align*}
parametrized by probability $\eta_i$ and decision threshold $\tau_i$. To further simplify, note that $Y_i$ is independent of $(X_{-i}, M)$ given $(X_i, S_i)$, i.e., a Markov chain
\begin{align*}
	(X_{-i}, M) - (X_i, S_i) - Y_i
\end{align*}
holds. Therefore,
\begin{align*}
	\LR(y_i | x^n, m) = \LR(y_i | x_i),
\end{align*}
and the equivalent detection rule is
\begin{align*}
	\Sh_i = \begin{cases}
		1 & \text{if } \LR(y_i | x_i) > \tau_i, \\
		1 \text{ with probability } \eta_i & \text{if } \LR(y_i | x_i) = \tau_i, \\
		0 & \text{if } \LR(y_i | x_i) < \tau_i.
	\end{cases}
\end{align*}
Note that the detection rule is parametrized by $(\eta_i, \tau_i)$, meaning that it depends on both $x_i$ and the time index $i$. In the following, we will show that it is sufficient to use time-independent detection rules.

For each $x$, let $\Ic_x$ be the set of time index $i$ such that $X_i = x$. Consider a subproblem
\begin{align*}
	\max &\sum_{i \in \Ic_x} P_{\D,i} \\
	\text{subject to } ~ &\sum_{i \in \Ic_x} P_{\FA,i} \le \alpha',
\end{align*}
and suppose that the solution $(P_{\FA,i}^*, P_{\D,i}^*)$ is attained by possibly distinct detection rules $\{\hat{S}_i\}_{i \in \Ic_x}$. Note that  $(P_{\FA,i}^*, P_{\D,i}^*)$ should be all boundary points on the receiver operating characteristic (ROC) curve for $x$. Otherwise, we can further increase the objective without increasing false alarm. Then, by the concavity of the ROC curves, there exists a single point $(P_{\FA,x}^\circ, P_{\D,x}^\circ)$ on the ROC curve such that
\begin{align*}
	\sum_{i \in \Ic_x} P_{\D,i}^* \le \sum_{i \in \Ic_x} P_{\D,x}^\circ = |\Ic_x| \cdot P_{\D,x}^\circ ~~ \text{ and } ~~ \sum_{i \in \Ic_x} P_{\FA,x}^\circ \le \alpha'.
\end{align*}
It in turn implies that a single (possibly randomized) decision rule is sufficient to solve the subproblem. Repeating the same argument for every $x$, we can conclude that it is sufficient to consider only $x$-dependent detection rules, which we have used in the achievability. Therefore, for any general detection rules $(\Sh_1, \ldots, \Sh_n)$, one can find $\{\Sh_x\}_{x \in \Xc}$ achieving equal or greater sum probability of detection at the same false alarm constraint. 

Recall that every codeword in $\Cc^{(n)}$ satisfies \eqref{eq:alpha_constraint}. It in turn implies that
\begin{align*}
	\alpha + \epsilon &\ge \max_{x^n \in \Cc^{(n)}} \frac{1}{n} \sum_{i=1}^n P^{(n)}_{\FA,i}(x^n) \\
	&\ge \max_{x^n \in \Cc_{p_X}^{(n)}} \frac{1}{n} \sum_{i=1}^n P^{(n)}_{\FA,i}(x^n) \\
	&\stackrel{(a)}{\ge} \frac{1}{n} \sum_{i=1}^n \sum_{x \in \Xc} \mathbf{1}\{ X_i = x \} \P[ \hat{S}_{x}(Y) = 1 | X = x, S = 0 ] \\
	&= \sum_{x} p_X(x) P_{\FA,x}(\Sh_x),
\end{align*}
where (a) follows since all codewords in $\Cc_{p_X}^{(n)}$ are of equal composition $p_X$, and time-independent decision rules achieving equal or smaller sum probability of false alarm can be found. Similarly, \eqref{eq:beta_constraint} implies that
\begin{align*}
	\beta-\epsilon &\le \min_{x^n \in \Cc^{(n)}} \frac{1}{n} \sum_{i=1}^n P^{(n)}_{\D,i}(x^n) \\
	&\le \min_{x^n \in \Cc_{p_X}^{(n)}} \frac{1}{n} \sum_{i=1}^n P^{(n)}_{\D,i}(x^n) \\
	&\le \frac{1}{n} \sum_{i=1}^n \sum_{x \in \Xc} \mathbf{1}\{ X_i = x \} \P[ \hat{S}_{x}(Y) = 1 | X = x, S = 1 ] \\
	&= \sum_{x} p_X(x) P_{\D,x}(\Sh_x).
\end{align*}
It concludes the converse proof.

\subsection{Numerical Examples}
\textbf{Binary Channel:} Consider binary state-dependent communication and sensing channels such that
\begin{align*}
	\Yt_i &= S_i x_i \oplus \Zt_i, ~ \Zt_i \sim \text{Bern}(\gamma_1), \\
	Y_i &= S_i x_i \oplus Z_i, ~ Z_i \sim \text{Bern}(\gamma_2),
\end{align*}
where $\Zt_i, Z_i$ are i.i.d.~Bernoulli noise with probability $\gamma_1, \gamma_2 < 0.5$, respectively. Also, assume that $S_i \sim \text{Bern}(\gamma_s)$ and $X_i \in \{0,1\}$.

When a constant composition code with composition $t := p_X(1)$ is used, the rate can be computed as follows. Noting that
\begin{align*}
	p_{\Yt}(1) &= (1-\gamma_s)\gamma_1 + \gamma_s ((1-t)\gamma_1 + t(1-\gamma_1)),
\end{align*}
the rate is bounded as
\begin{align*}
	R &\le I(p_X, p_{\Yt|X}) \\
	&= H_2(p_{\Yt}(1)) - (1-t)H(\Yt|X=0) - tH(\Yt|X=1) \\
	&= H_2(p_{\Yt}(1)) - (1-t)H_2(\gamma_1) \\
	&~~~~~~~~~~~~~~~~~~~~~~ - tH_2(\gamma_s(1-\gamma_1)+(1-\gamma_s)\gamma_1).
\end{align*}

For detection, note that when $X_i=0$, $Y_i=Z_i$ regardless of the state, i.e., the detector only observes pure noise. Therefore, it is optimal to make a random guess for $S$, and then, the ROC curve is indeed a straight line. On the other hand, when $X_i=1$, $Y_i=S_i \oplus Z_i$, i.e.,
\begin{align*}
	&H_0: Y_i \sim \text{Bern}(\gamma_2), \\
	&H_1: Y_i \sim \text{Bern}(1-\gamma_2).
\end{align*}
The ROC curve for these likelihoods is piecewise linear because the likelihood ratio can only take a finite number of values. Vertices of the ROC curve are achieved using deterministic tests, while line segments are achieved by randomizing deterministic tests achieving vertices.

\begin{figure}[t]
	\centering
	\subfloat[ROC curve for $X=1$]{\includegraphics[width=3.0in]{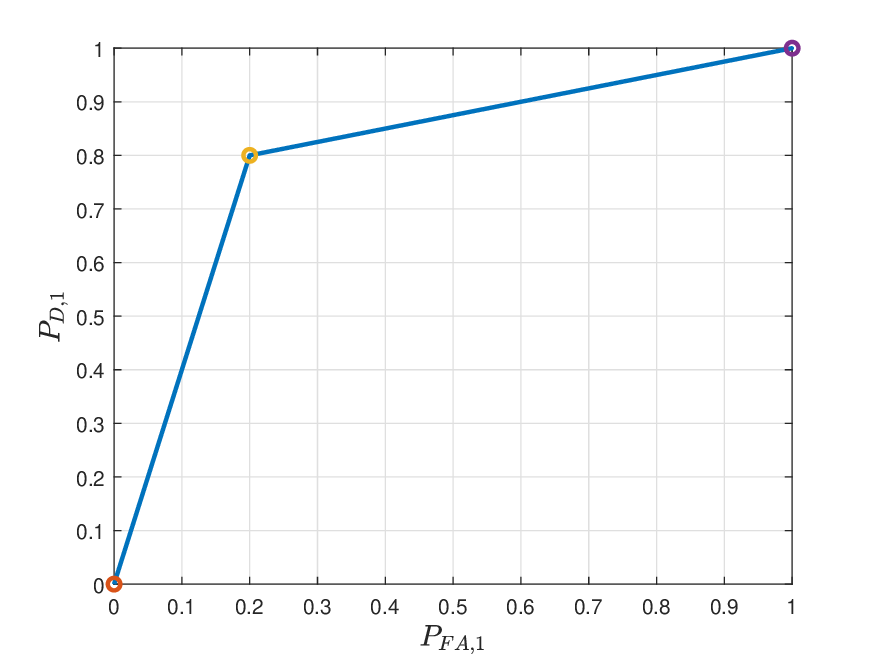}\label{fig:a_roc}} \\
	\subfloat[Rate-probability regions]{\includegraphics[width=3.0in]{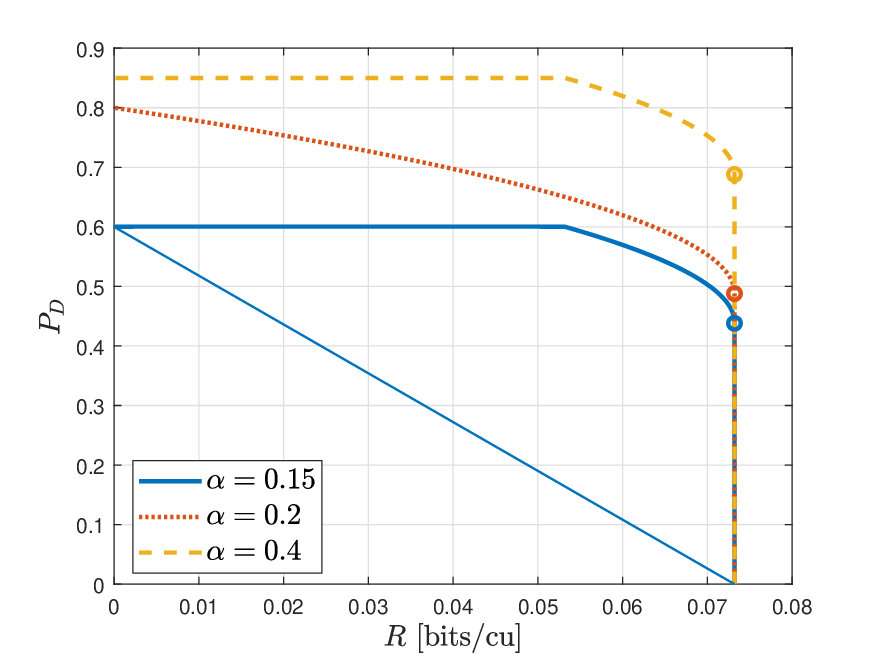}\label{fig:b_rate_prob}}
	\caption{Binary example assuming $\gamma_1 = \gamma_2 = 0.2$. In (b), the maximal $P_\D$ achieved at the highest data rate (i.e., channel capacity) is indicated by circle, corresponding to the input distribution $p_X(1) = 0.4824$.}
	\label{fig:binary_iid_ex}
\end{figure}

For numerical evaluation, consider the case where $\gamma_s = 0.5$ and $\gamma_1 = \gamma_2 = 0.2$. Figure \ref{fig:a_roc} shows the ROC curve for $X=1$, which is piecewise linear and concave as stated. The lower-left circle represents the case where $\Sh_1(y) = 0$ for all $y$, while the upper-right circle represents the case where $\Sh_1(y) = 1$ for all $y$. The middle circle is obtained by declaring $\Sh_1(y) = y$, which gives
\begin{align*}
	P_{\FA, 1} &= \P_0[ \Sh_1(Y)=1 | X=1] \\
    &= \P_0[ Y=1 | X=1 ] = \gamma_2, \\
	P_{\D, 1} &= \P_1[ \Sh_1(Y)=1 | X=1 ] \\
    &= \P_1[ Y=1 | X=1 ] = 1-\gamma_2.
\end{align*}
The line segments are attained by randomizing two tests for vertices. The slope for the left segment is $\frac{1-\gamma_2}{\gamma_2} > 1$ and that for the right segment is $\frac{\gamma_2}{1-\gamma_2} < 1$.

Figure \ref{fig:b_rate_prob} illustrates the cross sections of rate-probability region at $\alpha = 0.15, 0.2, 0.4$. The circles indicate points at which the channel capacity is achieved. As expected, a larger $\alpha$ results in a larger region. Additionally, it is interestingly observed that the maximum probability of detection tends to remain flat at low data rates and decreases smoothly until the data rate reaches the channel capacity. This flat behavior is attributed to the (piecewise) linear nature of the ROC curves. To see this, since $P_{\FA,0} = P_{\D,0}$ for $X=0$, first note that the optimization problem is
\begin{align*}
	\max ~ &(1-p(1)) P_{\FA,0} + p(1) P_{\D,1} \\
	\text{subject to } &(1-p(1)) P_{\FA,0} + p(1) P_{\FA,1} \le \alpha.
\end{align*}
Note also that when $X=0$, the slope (i.e., the gain in the objective) of the ROC is $1$. When $X=1$, the ROC curve is piecewise linear, with an initial slope $\frac{1-\gamma_2}{\gamma_2} > 1$ and a final slope $\frac{\gamma_2}{1-\gamma_2} < 1$. Hence, as $\alpha$ increases, the false alarm budget is allocated in the decreasing order of slope. Specifically, the false alarm budget is first assigned to $P_{\FA,1}$ until it reaches $P_{\FA,1} = \gamma_2$, where the slope is largest. Next, the budget is allocated to $P_{\FA,0}$ until it reaches $P_{\FA,0} = 1$. Finally, any remaining budget is assigned to $P_{\FA,1}$, where the slope is the smallest. Considering each case separately, we can prove the flat behavior of the region. When $\alpha$ is small (e.g., $\alpha = 0.15$) and $t$ exceeds a certain threshold, increasing $P_{\FA,1}$ offers a greater gain of $\frac{1 - \gamma_2}{\gamma_2}$ compared to increasing $P_{\FA,0}$. This means that $P_{\FA,0}$ is set to $0$, and the entire false alarm budget is allocated to $P_{\FA,1}$. Solving the optimization problem with $P_{\D,1} = \frac{1 - \gamma_2}{\gamma_2} P_{\FA,1}$ shows that the objective function remains constant. Similarly, when $\alpha$ is large (e.g., $\alpha = 0.5$) and $t$ exceeds a certain threshold, the large false alarm budget means the optimal solution occurs when $P_{\FA,0} = 1$ and $P_{\FA,1} \in (\gamma_2, 1]$. In this case, $P_{\D,1} = \frac{\gamma_2}{1 - \gamma_2} (P_{\FA,1} - \gamma_2) + (1 - \gamma_2)$, and solving the optimization problem shows that the objective function is also constant. For $\alpha = 0.2$, the situation is between these two extremes. When the optimal values are $P_{\FA,0} \in (0,1)$ and $P_{\FA,1} = \gamma_2$, solving the optimization problem results in a $t$-dependent objective function. This indicates a non-flat tradeoff due to the coupling of the probability of detection and data rate by $p_X$. For this example, a na\"{i}ve time sharing strategy achieves the linear line connecting the two extreme points; see the solid linear line in Fig.~\ref{fig:b_rate_prob}. Other time sharing points are omitted for visual clarity.

\textbf{Gaussian Channel:} Next, consider an example involving a Gaussian channel with on-off signaling. Suppose that the state changes channel gain, i.e.,
\begin{align*}
	\Yt_i &= S_i x_i + \Zt_i, ~~~ \Zt_i \sim \Nc(0, \sigma_{c}^2), \\
	Y_i &= S_i x_i + Z_i, ~~~ Z_i \sim \Nc(0, \sigma_{s}^2).
\end{align*}
Also, suppose that $S_i \sim \text{Bern}(\gamma_s)$ and $X \in \{0, 1\}$. Then, letting $t := p_X(1)$,
\begin{align*}
	\Yt &\sim (1-\gamma_s)\Nc(0, \sigma_c^2) + \gamma_s (1-t) \Nc(0, \sigma_c^2) + \gamma_s t \Nc(1, \sigma_c^2),
\end{align*}
and
\begin{align*}
	\Yt|X=0 &\sim \Nc(0, \sigma_{c}^2), \\
	\Yt|X=1 &\sim (1-\gamma_s) \Nc(0, \sigma_{c}^2) + \gamma_s \Nc(1, \sigma_{c}^2).
\end{align*}
Then, the data rate is
\begin{align*}
	R &< I(X;\Yt) \\
	&= h(\Yt) - (1-t)h(\Yt | X=0) - t \cdot h(\Yt | X=1) \\
	&= h(\Yt) - \frac{1-t}{2} \log (2\pi e \sigma_c^2) - t \cdot h(\Yt | X=1).
\end{align*}

For detection, note that $\Yt_i = \Zt_i$ when $X_i=0$. Hence, the ROC curve for $X=0$ is a straight line with slope $1$. When $X_i=1$, the likelihoods are
\begin{align*}
	Y_i &\sim \Nc(0, \sigma_s^2) ~~ \text{if } S_i=0, \\
	Y_i &\sim \Nc(1, \sigma_s^2) ~~ \text{if } S_i=1,
\end{align*}
which implies the likelihood ratio test such that
\begin{align*}
	\frac{ \frac{1}{\sqrt{2\pi \sigma_s^2}} \exp\left( -\frac{(y-1)^2}{2 \sigma_s^2}\right) }{ \frac{1}{\sqrt{2\pi \sigma_s^2}} \exp\left( -\frac{y^2}{2 \sigma_s^2}\right) } > \tau ~ \Leftrightarrow ~ y > \frac{ 2\sigma_s^2 \log \tau + 1 }{2} =: \tau'.
\end{align*}
Hence, the ROC curve can be written in closed form using the $Q$-function as follows:
\begin{align*}
	P_{\FA,1} &= \P_0[ Y > \tau' | X=1 ] \\
    &= Q( \tau'' ), \\
	P_{\D,1} &= \P_1[ Y > \tau' | X=1 ] \\
    &= Q \left( \tau'' - \frac{1}{\sigma_s} \right),
\end{align*}
where $\tau'' := \frac{\tau'}{\sigma_s} = \frac{ 2\sigma_s^2 \log \tau + 1 }{2 \sigma_s}$. Furthermore, using the fact that
\begin{align*}
	\frac{d Q(z)}{d z} = - \frac{1}{\sqrt{2\pi}} \exp\left( -\frac{z^2}{2} \right),
\end{align*}
the slope of the ROC curve at $\tau''$ is
\begin{align*}
	\frac{d P_{\D,1}}{d P_{\FA,1}} &= \frac{d P_{\D,1}}{d z} \cdot \frac{d z}{d P_{\FA,1}} \bigg\vert_{z = \tau''} \\
	&= \exp \left( -\frac{(\tau''- 1/\sigma_s)^2}{2} + \frac{(\tau'')^2}{2} \right) = \tau,
\end{align*}
which is the same as the decision threshold we used. This is generally true that $\frac{d P_{\D,1}}{d P_{\FA,1}}$ at a point on the ROC curve is the same as the threshold of the likelihood ratio test that achieves the $P_{D,1}$ and $P_{\FA,1}$ at that point~\cite{MoulinV2019}.

\begin{figure}[t]
	\begin{center}
		\includegraphics[width=3.0in]{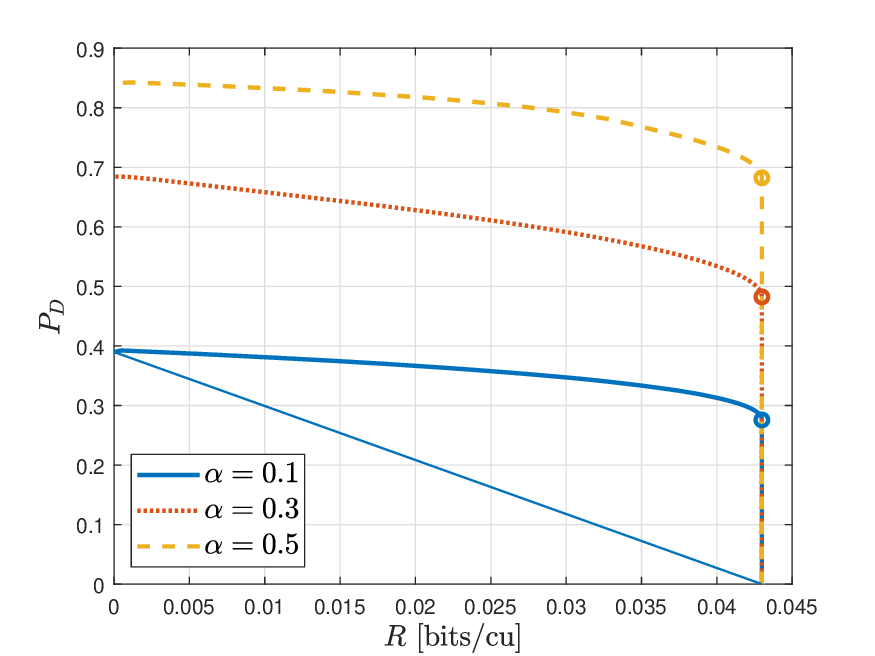}
	\end{center}
	\caption{Rate-probability region for the Gaussian example. $\gamma_s = 0.5$ and $\sigma_c^2 = \sigma_s^2 = 1$ are assumed, and $\alpha=0.1, 0.3, 0.5$ are plotted. Maximal $P_\D$ achieved with the maximal data rate (i.e., channel capacity) is marked by circle, corresponding to the input distribution $p_X(1) = 0.4760$.}
	\label{fig:gaussian_iid_ex}
\end{figure}

Consider a specific example where $\sigma_c^2 = \sigma_s^2 = 1$ and $\gamma_s = 0.5$. The rate-probability regions for $\alpha=0.1, 0.3, 0.5$ are shown in Fig.~\ref{fig:gaussian_iid_ex}, where we evaluated the mutual information through numerical integration. Unlike the binary example, the boundary strictly decreases as the rate increases, until it reaches the capacity of the communication channel. Similar to the binary example, a na\"{i}ve time sharing strategy achieves the linear line connecting the two extreme points; see the solid linear line in Fig.~\ref{fig:gaussian_iid_ex}. Other time sharing cases are omitted for visual clarity. 

\section{Conclusion}\label{sec:discussion}
In this paper, we characterize the optimal three-way tradeoff between coding rate and probabilities of false alarm and missed detection. For the fixed state setting, we introduced Hoeffding's hypothesis testing problem~\cite{Hoeffding1965} into the ISAC framework and extended prior literature focusing on average error constraints~\cite{Joudeh--Willems2022, Chang--Wang--Erdogan--Bloch--2023} to the more general scenario of unequal error constraints. In the i.i.d.~state setting, we extend the work in \cite{Ahmadipour--Kobayashi--Wigger--Caire2022} to unequal error constraints, offering a more nuanced understanding of error events. Our results provide new insights into how to simultaneously detect and manage the two types of error events with an information-bearing signal, presenting the theoretical utility and practical significance of these approaches.

While our study provides a theoretical foundation, several directions considering relevant practical aspects remain. For instance, extending the framework to multi-target detection or multi-user communication scenarios could address broader real-world challenges. Incorporating interference models, such as inter-symbol interference or multi-user interference, would further enhance the practical applicability of the results. Exploring the impact of new technologies like reconfigurable intelligent surfaces (RIS) within this framework also represents a promising avenue for further research.
These potential extensions, while not the primary focus here, highlight the versatility of the developed theoretical framework and its potential to develop future advancements in ISAC technologies.

\appendices

\section{Bound on $\P(T_v=1|x^n)$}\label{app:appA}
Recall that $\tau = \E_v\left[ \frac{1}{n} \LLR(Y^n|x^n) \right] - \delta$ where $\delta > 0$. Then,
\begin{align*}
	\P( T_v=0 | x^n ) &= \sum_{y^n: \LLR(y^n|x^n) < n\tau}p_v( y^n|x^n)   \\
	&= \sum_{y^n: \LLR(y^n|x^n)-n\mu_v < -n\delta}p_v( y^n|x^n)   \\
	&\le\sum_{y^n: |\LLR(y^n|x^n)-n\mu_v| > n\delta}p_v( y^n|x^n)\\
	&\stackrel{(a)}{\le} \frac{ \text{Var}(\LLR(Y^n|x^n)) }{ n^2\delta^2} \\
	&\stackrel{(b)}{\le} \frac{C}{ n\delta^2},
\end{align*}
where $(a)$ follows from the Chebyshev inequality, and $(b)$ follows from our assumption~\eqref{eq:assupmt1}. Since $\P(T_v=0|x^n) = 1-\P(T_v=1|x^n)$, it holds that
$\P(T_v=1|x^n) \ge 1 - \frac{C}{n \delta^2}.$


\begin{thebibliography}{99}	
\bibitem{Bourdoux2020_2}
A.~{Bourdoux \textit{et al.}}, ``6{G} white paper on localization and
sensing,'' 2020, preprint available at https://arxiv.org/abs/2006.01779.

\bibitem{Liu--Huang--Li--wan--Li--Han--Liu--Du--Tan--Lu--Shen--Colone--Chetty2022_2}
A.~{Liu \textit{et al.}}, ``A survey on fundamental limits of integrated
sensing and communication,'' \emph{{IEEE} Commun. Surveys Tuts.}, vol.~24,
no.~2, pp. 994--1034, 2022.

\bibitem{Liu--Masouros--Petropulu--Griffiths--Hanzo2020}
F.~Liu, C.~Masouros, A.~P. Petropulu, H.~Griffiths, and L.~Hanzo, ``Joint radar
and communication design: {A}pplications, state-of-the-art, and the road
ahead,'' \emph{{IEEE} Trans. Commun.}, vol.~68, no.~6, pp. 3834--3862, Jun.
2020.

\bibitem{Sturm--Wiesbeck2011}
C.~Sturm and W.~Wiesbeck, ``Waveform design and signal processing aspects for
fusion of wireless communications and radar sensing,'' \emph{Proceedings of
	the IEEE}, vol.~99, no.~7, pp. 1236--1259, Jul. 2011.

\bibitem{Joudeh--Willems2022}
H.~Joudeh and F.~M.~J. Willems, ``Joint communication and binary state
detection,'' \emph{{IEEE} J. Sel. Areas Inf. Theory}, vol.~3, no.~1, pp.
113--124, Mar. 2022.

\bibitem{Wu--Joudeh2022}
H.~Wu and H.~Joudeh, ``On joint communication and channel discrimination,'' in
\emph{Proc. 2022 IEEE Int. Symp. Inf. Theory}, June-July 2022, pp.
3321--3326.

\bibitem{Chang--Wang--Erdogan--Bloch--2023}
M.-C. Chang, S.-Y. Wang, T.~Erdoğan, and M.~R. Bloch, ``Rate and
detection-error exponent tradeoff for joint communication and sensing of
fixed channel states,'' \emph{{IEEE} J. Sel. Areas Inf. Theory}, vol.~4, pp.
245--259, May 2023.

\bibitem{Ahmadipour--Kobayashi--Wigger--Caire2022}
M.~Ahmadipour, M.~Kobayashi, M.~Wigger, and G.~Caire, ``An
information-theoretic approach to joint sensing and communication,''
\emph{{IEEE} Trans. Inf. Theory}, vol.~70, no.~2, pp. 1124--1146, Feb. 2024.

\bibitem{Poor1994}
H.~V. Poor, \emph{An Introduction to Signal Detection and Estimation}.\hskip
1em plus 0.5em minus 0.4em\relax New York, USA: Springer-Verlag, 1994.

\bibitem{MoulinV2019}
P.~Moulin and V.~Veeravalli, \emph{Statistical Inference for Engineers and Data
	Scientists}.\hskip 1em plus 0.5em minus 0.4em\relax Cambridge, U.K.:
Cambridge University Press, 2019.

\bibitem{AhmadipourWS2024}
M.~Ahmadipour, M.~Wigger, and S.~Shamai, ``Strong converse for bi-static isac
with two detection-error exponents,'' in \emph{Proc. 2024 Int. Zurich Seminar
	Inf. Theory (IZS'24)}, Mar. 2024.

\bibitem{Zhang--Vedantam--Mitra2011}
W.~Zhang, S.~Vedantam, and U.~Mitra, ``Joint transmission and state estimation:
{A} constrained channel coding approach,'' \emph{{IEEE} Trans. Inf. Theory},
vol.~57, no.~10, pp. 7084--7095, Oct. 2011.

\bibitem{Sutivong--Chiang--Cover--Kim2005}
A.~Sutivong, M.~Chiang, T.~M. Cover, and Y.-H. Kim, ``Channel capacity and
state estimation for state-dependent {G}aussian channels,'' \emph{{IEEE}
	Trans. Inf. Theory}, vol.~51, no.~4, pp. 1486--1495, Apr. 2005.

\bibitem{Kim--Sutivon--Cover2008}
Y.-H. Kim, A.~Sutivong, and T.~M. Cover, ``State amplification,'' \emph{{IEEE}
	Trans. Inf. Theory}, vol.~54, no.~5, pp. 1850--1859, May 2008.

\bibitem{Choudhuri--Kim--Mitra2020}
C.~Choudhuri, Y.-H. Kim, and U.~Mitra, ``Causal state amplification,'' in
\emph{Proc. 2011 IEEE Int. Symp. Inf. Theory}, July-Aug. 2011, pp.
2110--2114.

\bibitem{LiADMB2024}
X.~Li, V.~C. Andrei, A.~Djuhera, U.~J. M\"{o}nich, and H.~Boche, ``An analysis
of capacity-distortion trade-offs in memoryless isac systems,'' 2024,
preprint available at https://arxiv.org/pdf/2402.17058.

\bibitem{GootyM2024}
S.~A. Gooty and H.~Mahdavifar, ``Strategies for rate optimization in joint
communication and sensing over channels with memory,'' in \emph{Proc. 60th
	Annu. Allerton Conf. Commun. Control Comput.}, Sep. 2024.

\bibitem{Xiong--Liu--Cui--Yuan--Han--Caire2023}
Y.~Xiong, F.~Liu, Y.~Cui, W.~Yuan, T.~X. Han, and G.~Caire, ``On the
fundamental tradeoff of integrated sensing and communications under
{G}aussian channels,'' \emph{{IEEE} Trans. Inf. Theory}, vol.~69, no.~9, Sep.
2023.

\bibitem{Liu--Yuan--Masouros--Yuan2020}
F.~Liu, W.~Yuan, C.~Masouros, and J.~Yuan, ``Radar-assisted predictive
beamforming for vehicular links: {C}ommunication served by sensing,''
\emph{{IEEE} Trans. Wireless Commun.}, vol.~19, no.~11, pp. 7704--7719, Nov.
2020.

\bibitem{Liu--Liu--Li--Masouros--Eldar2022}
F.~Liu, Y.-F. Liu, A.~Li, C.~Masouros, and Y.~C. Eldar, ``Cram\'{e}r-rao bound
optimization for joint radar-communication beamforming,'' \emph{{IEEE} Trans.
	Signal Process.}, vol.~70, pp. 240--253, Feb. 2022.

\bibitem{Hua--Han--Xu2023}
H.~Hua, T.~X. Han, and J.~Xu, ``{MIMO} integrated sensing and communication:
{CRB}-rate tradeoff,'' \emph{{IEEE} Trans. Wireless Commun.}, vol.~23, no.~4,
pp. 2839--2854, Apr. 2024.

\bibitem{DongLLX2023}
F.~Dong, F.~Liu, S.~Lu, and Y.~Xiong, ``Rethinking estimation rate for wireless
sensing: A rate-distortion perspective,'' \emph{{IEEE} Trans. Veh. Technol.},
vol.~72, no.~12, pp. 16\,876--16\,881, Dec. 2023.

\bibitem{RenPSFQLNX2024}
Z.~Ren, Y.~Peng, X.~Song, Y.~Fang, L.~Qiu, L.~Liu, D.~W.~K. Ng, and J.~Xu,
``Fundamental {CRB}-rate tradeoff in multi-antenna {ISAC} systems with
information multicasting and multi-target sensing,'' \emph{{IEEE} Trans.
	Wireless Commun.}, vol.~23, no.~4, pp. 3870--3885, Apr. 2024.

\bibitem{Kobayashi--Hamad--Kramer--Caire2019}
M.~Kobayashi, H.~Hamad, G.~Kramer, and G.~Caire, ``Joint state sensing and
communication over memoryless multiple access channels,'' in \emph{Proc. 2019
	IEEE Int. Symp. Inf. Theory}, Jul. 2019, pp. 270--274.

\bibitem{Liu--Li--Ong--Yener2024}
Y.~Liu, M.~Li, L.~Ong, and A.~Yener, ``Bistatic integrated sensing and
communication over memoryless relay channels,'' in \emph{Proc. 2024 IEEE Int.
	Symp. Inf. Theory}, Jul. 2024, pp. 2592--2597.

\bibitem{Hoeffding1965}
W.~Hoeffding, ``Asymptotically optimal tests for multinomial distributions,''
\emph{Ann. Math. Stat.}, vol.~36, no.~2, pp. 369--401, Apr. 1965.

\bibitem{Wu--Joudeh2024}
H.~Wu and H.~Joudeh, ``Joint communication and channel discrimination,''
\emph{Entropy}, vol.~26, no.~12, 2024.

\bibitem{Seo--Lim_wiopt2024}
D.~Seo and S.~H. Lim, ``Integrated communication and binary state detection
from {H}oeffding's perspective,'' in \emph{Proc. 22nd Int. Symp. Model.
	Optim. Mobile, Ad Hoc, Wireless Netw. (WiOpt '24)}, Oct. 2024, to appear.

\bibitem{El-Gamal--Kim2011}
A.~El~Gamal and Y.-H. Kim, \emph{Network Information Theory}.\hskip 1em plus
0.5em minus 0.4em\relax Cambridge: Cambridge University Press, 2011.

\bibitem{CoverT1991}
T.~M. Cover and J.~A. Thomas, \emph{Elements of Information Theory}.\hskip 1em
plus 0.5em minus 0.4em\relax New York, USA: John Wiley \& Sons, 1991.

\bibitem{Csiszar--Korner2011}
I.~{Csisz\'{a}r} and J.~{K\"{o}rner}, \emph{Information Theory: Coding Theorems
	for Discrete Memoryless Systems}, 2nd~ed.\hskip 1em plus 0.5em minus
0.4em\relax Cambridge, U.K.: Cambridge Univ. Press, 2011.

\bibitem{Telatar1999}
{\.I}.~E. Telatar, ``Capacity of multi-antenna {G}aussian channels,''
\emph{Euro. Trans. Telecomm.}, vol.~10, no.~6, pp. 585--595, Nov.-Dec. 1999.

\bibitem{BoydV2004}
S.~Boyd and L.~Vandenberghe, \emph{Convex optimization}.\hskip 1em plus 0.5em
minus 0.4em\relax Cambridge, U.K.: Cambridge university press, 2004.
\end{thebibliography}
\end{document}